# Nonlocal fractional Kardar-Parisi-Zhang dynamics of grain boundaries

Kai Zhao[1,*]

*1. School of Mechanical Engineering, Jiangnan University, Wuxi 214122, China*

The kinetic roughening of driven grain boundaries (GBs) is mediated by the collective behavior of disconnections, whose long-range $1/r$ elastic interactions distinguish them from conventional growing surfaces. Large-scale molecular dynamics simulations show that the roughness of driven GBs undergoes a sharp crossover as the applied strain increases, with the Hurst exponent ($H$) increasing from approximately 0.33 to 0.8. We develop a nonlocal fractional Kardar-Parisi-Zhang (fKPZ) theory to describe this low-temperature behavior, where the $1/r$ nonlocal elasticity behaves as the fractional relaxation of the GB morphology. The results show that the spatially-correlated noise associated with disconnection avalanches interacts with the KPZ nonlinearity to produce the post-crossover roughening observed in atomistic simulations. The post-crossover roughness exponent ($H \cong 0.8$) is comparable to values reported for dynamic fracture, suggesting the underlying physics shared by the collective behavior of defects (e.g., disconnections, dislocations) and interfacial roughening.

## 1. Introduction

The migration of grain boundaries (GBs) governs the microstructural evolution, plastic deformation, and macroscopically thermomechanical properties of polycrystalline materials [1-5]. Far from the classical picture of continuous, curvature-driven flow, GB migration is now established to be mediated by the nucleation and glide of discrete topological line defects known as disconnections [5-7]. Each disconnection is characterized by a dual topological invariant, *i.e.*, the Burgers vector $\boldsymbol{b}$ quantifying the interfacial shear displacement, and the step height $\boldsymbol{h}$ quantifying the normal migration, that together provide the mechanistic basis for shear-coupled GB migration [8, 9]. Over the past decades, extensive theoretical and computational efforts have been dedicated to formulate continuous equations of motion for GBs based on multi-disconnection-mode kinetics [2, 6, 10, 11]. These models have been successful in explaining numerous nonclassical behaviors, including the non-Arrhenius and cryogenic migration kinetics [12, 13], the energetic pathways and nucleation barriers of specific disconnection modes [7], disconnection-mediated sliding in diverse crystal structures [14], and topological lattice transformations during migration [15]. Furthermore, recent atomistic studies have revealed even richer sub-GB dynamics, such as multi-disconnection decomposition and composition under loading [16], as well as the critical role of complexion in driving GB phase transitions in metals [17, 18].

* Corresponding author: kai.zhao@jiangnan.edu.cn (K. Z.)

Despite the success of disconnection theory in unifying the GB behavior, a fundamental gap persists. The vast majority of current analytical and mesoscale models implicitly assume that the GB remains macroscopically flat or adopt an idealized, smoothly-curved profile during migration. They neglect the complex, far-from-equilibrium evolution of the morphology, *i.e.*, the kinetic roughening, that occurs when the interface is subjected to severe mechanical loading [3, 19-21]. In reality, plastically-deformed interfaces are significantly corrugated by the structural relaxation and defect pile-ups locally, thus, the flat-GB assumption breaks down in predicting the actual rheology of polycrystals.

To accurately describe the non-planar, dynamic morphology of evolving interfaces, one has to invoke the framework of nonequilibrium statistical mechanics and fractal geometry. It is widely recognized that driven interfaces in complex systems develop the scale-invariant, self-affine roughness characterized by the Hurst exponent $H$ [19, 22, 23]. The Edwards-Wilkinson (EW) equation describes the relaxation of stochastically driven interfaces with short-range elastic restoring forces, while its nonlinear generalization, the Kardar-Parisi-Zhang (KPZ) equation, accounts for the lateral growth that alters the universality class [24, 25]. In the context of GBs, thermal fluctuations of shear-coupled interfaces at equilibrium are constrained by long-range elastic interactions, yielding a specific spatial fluctuation spectrum [1] and corresponding random-walk mobilities extracted via the fluctuation-dissipation theorem (FDT) [3]. However, real GBs under applied strain rates operate far from thermal equilibrium. Experiments and atomistic simulations have demonstrated that self-affine fractal surfaces can be produced by plastic deformation, with the roughness exponent evolves with strain dynamically [26, 27]. Besides, driven systems with nonlocal interactions or structural disorder can exhibit kinetic super-roughening and anomalous dynamic scaling, where the global roughness exponent significantly exceeds that of the corresponding short-range universality class [28, 29]. Moreover, the structural rearrangements evidenced by the string-like collective motion and transient caging during plastic flow are observed in glass-forming liquids and dense active matter [30]. However, a unified theory bridging the long-range elastic interactions of disconnections with the interfacial roughening and dynamic phase behavior of GBs remains outstanding.

In this work, we bridge the gap between micro-mechanics of disconnections and non-equilibrium statistical physics of GBs. The dynamic roughening transition of shear-coupled GBs is studied using large-scale molecular dynamics (MD) simulations and the fractional KPZ (fKPZ) framework together. We demonstrate that the onset of GB morphology crossover is not merely the amplification of stochastic fluctuations, but also the dynamic depinning transition triggered by disconnection avalanches. Unlike impurity-pinned interfaces, the discrete step character and $1/r$ nonlocal elasticity of disconnections could introduce the corrugated energy landscape, giving rise to the depinning dynamics even in chemically-pure bicrystals. By representing the $1/r$ elasticity among disconnections as the *fractional Laplacian*, we obtain the fKPZ equation as the governing *stochastic partial differential equation* (SPDE) for the GB height profile. The theory reveals that the (deterministic) *fractional Burgers equation* describes the competition between nonlinear steepening and nonlocal elastic relaxation, while the collective motion of disconnections is represented by the spatially-correlated noise, contributing to the post-crossover roughening with a scaling plateau of the Hurst exponent $H \cong 0.8$. It is thus established that a transition of roughness scaling occurs, from the quenched Edwards-Wilkinson (qEW) to the fKPZ regime.

## 2. Transition of Hurst exponent

### 2.1. MD protocols

We perform MD simulations for an fcc Ni bicrystal containing a $\Sigma 5[100](310)$ symmetric tilt GB under constant strain-rate shear loading at $T$ = 100 ~ 1000 K. The strain rates, $\dot{\gamma}$ = $10^6$ and $10^7$ $s^{-1}$, are imposed by applying corresponding velocities in the boundary regions along the *x*-axis direction in the *xy*-plane, as shown in **Fig. 1**. A quasi-2D simulation box of ~ 111.7×335.2×0.35 nm is built to ensure GBs do not interfere with each other along the *y*-axis direction. Periodic boundary conditions are applied along the *x*- and *z*-axis directions. The interatomic interactions are described by using the embedded-atom method potential [31]. All MD simulations are performed by using the LAMMPS code [32]. The GB morphology is identified by using the common neighbor analysis and cluster analysis tools within the OVITO code [33]. Similar as previous study [34], **Fig. 1** shows that a constant velocity $v_u$ is applied in the upper boundary, while the lower boundary is fixed (*i.e.*, its velocity $v_l = 0$). As argued by Jian *et al*. [34], a 'partial thermostat' (similar to 'stadium BCs') is realized by applying an NVT ensemble to regions near boundaries and NVE ensemble to central region, which could maintain the desired temperature while not interfering with the motion of microstructures. For the Mishin potential [31] of fcc Ni, the elastic constants are, $C_{11}$ = 247.862 GPa, $C_{12}$ = 147.828 GPa, $C_{44}$ = 124.838 GPa. The rotation angle of the present bicrystalline specimen is $\theta' = \pm 18.43°$ (the misorientation angle is thus $\theta = 2|\theta'| = 36.87°$) [35]. The theoretical shear modulus is thus, $C'_{66} = C_{44} + \sin^2(\theta)\left[\frac{C_{11}-C_{12}}{2} - C_{44}\right]$. The calculated value is ~ 97.9 GPa, which would be compared with the MD results in **§2.2** later.

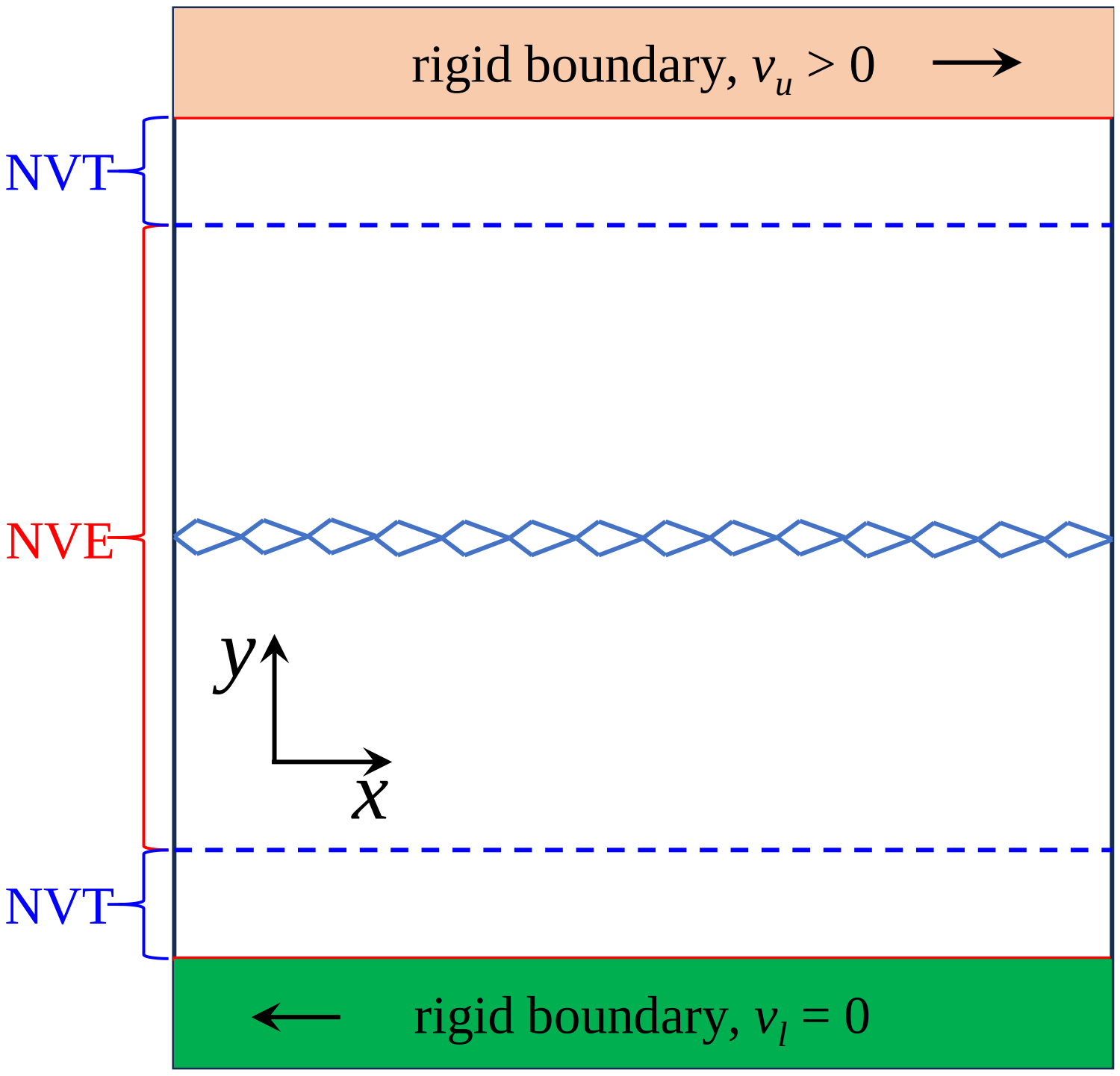

**Fig. 1**. Schematic of the bicrystalline specimen.

The GB morphology was extracted from the output atomic configurations with the following procedures, 1) ***common neighbor analysis*** was applied to identify atomic structure; 2) all fcc atoms were deleted; 3) ***cluster analysis*** was applied with a cutoff = 5.0 Å; 4) all clusters with the index larger than 1 were deleted; 5) remaining atoms were saved as new files for the power spectral density (PSD) computation. The GB morphologies at different temperatures and strain rates are shown in **Fig. S1** ~ **Fig. S7**, respectively (see **Supplementary Materials**), and indicate that the GB profile is roughened as the strain increases.

### 2.2 MD results

The instantaneous GB height profile $y(x,t)$ was tracked via cluster analysis, and the Hurst exponent $H$ was computed as a function of applied shear strain $\gamma$ through PSD analysis with $S(q) \propto q^{-(1+2H)}$. **Fig. S8** ~ **Fig. S14** (see **Supplementary Materials**) show the evolution of PSD curves under different temperatures and strain rates, respectively. It should be noted that a spacing factor ($C > 1$) is introduced in these figures for visualization, but does not affect the results, *i.e.*, $\Phi_{stacked}(q) = \Phi(q) \times C^{k-1}$, where, $k$ is the index of selected data trajectories, which takes the integer value in sequence within [1, $n$], where $n$ is the number of PSD curves to be displayed. The extracted evolutions of Hurst exponent $H$ against the shear strain $\gamma$ are shown in **Fig. 2** here.

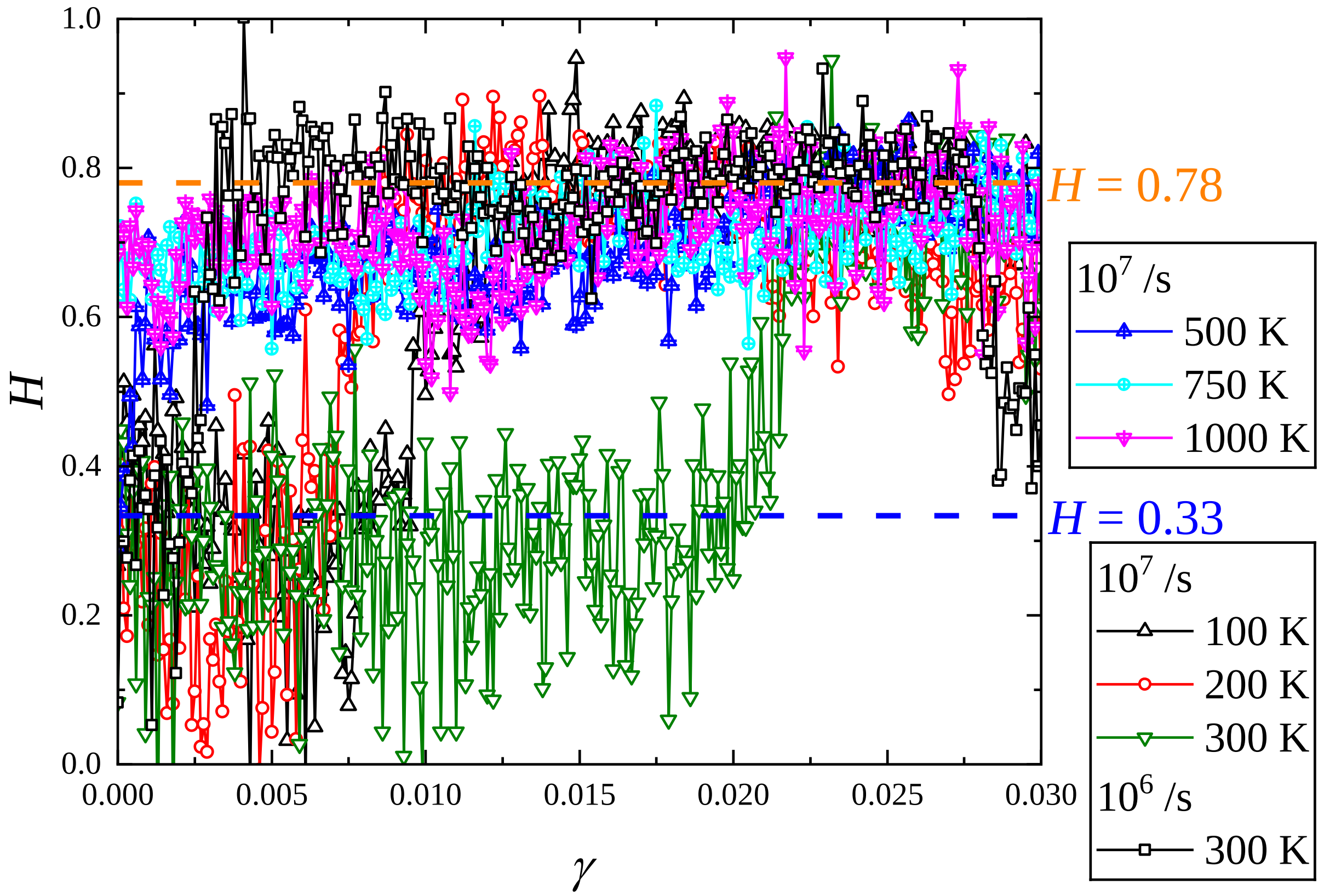

**Fig. 2**. Evolution of the Hurst exponent against the shear strain.

The evolution of the Hurst exponent $H$ with strain at $T$ = 100, 200 and 300 K reveals two features, *i.e.*,

*1) Pre-crossover regime*. At small strains, the Hurst exponent stabilizes at $H \cong 0.33$. This value lies below the uncorrelated random-walk limit ($H = 0.5$), demonstrating that $1/r$ elasticity suppresses stochastic fluctuations. This pre-crossover morphology is discussed below in terms of nonlocal elasticity and lattice pinning.

*2) Morphological crossover*. With increasing shear strain, the Hurst exponent rises sharply toward $H \cong 0.8$, indicating a change in the spatial roughness of the migrating GB. However, results at $T$ = 500, 750, and 1000 K (see **Fig. 2**) exhibit that the Hurst exponent already exceeds the $H = 0.5$ limit before the crossover, although they approach to the post-crossover value of $H \cong 0.8$ as the strain increases.

We further compare the stress-strain curves under different temperatures in **Fig. 3**, where the fitted shear modulus (of the 500 K curve) is about 99.89 GPa, in good agreement with the theoretical value ~ 97.9 GPa predicted by the anisotropic elasticity (see **§2.1**).

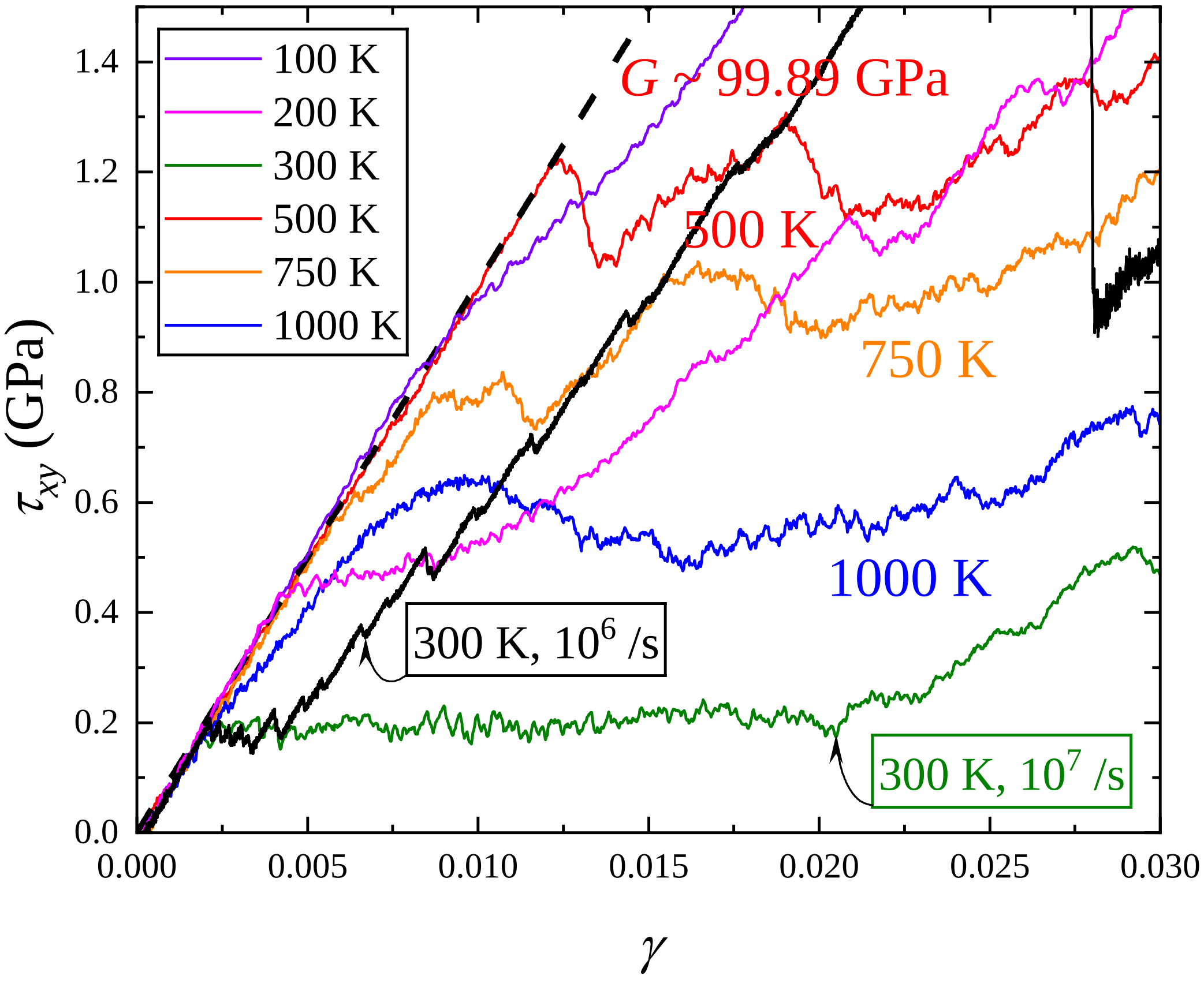

**Fig. 3**. Temperature-dependent stress-strain responses. Except the black solid curve is under $\dot{\gamma}$ = $10^6$ /s, other colored (solid) curves are under $\dot{\gamma}$ = $10^7$ /s. The black dash line represents the linear fit to the initial elastic branch of the 500 K curve under $\dot{\gamma}$ = $10^7$ /s, yielding an effective shear modulus of $G$ ~ 99.89 GPa, which is in excellent agreement with the prediction of anisotropic elasticity.

Besides, the extracted stresses exhibit the non-Arrhenius dependence of temperature [13, 36, 37], with the flow stress peaking at intermediate temperatures ($\tau_{500\,K} > \tau_{750\,K} > \tau_{1000\,K} > \tau_{300\,K}$) for $\dot{\gamma}$ = $10^7$ /s. This non-monotonic behavior indicates the interplay between the collective behavior of disconnections and thermal fluctuations. At low temperatures (100, 200 and 300 K in present study), the structurally ordered interface facilitates the facile, low-barrier glide of disconnections, necessitating the minimal stress. However, as temperature increases (500 K), the thermally roughened interface manifests as an array of dynamic pinning centers, retarding the disconnection motion and thereby demanding larger driving stress.

At even higher temperatures (750 K - 1000 K), abundant thermal activation (e.g., atomic shuffling) overrides the pinning landscape, inducing interfacial softening and recovering the classical Arrhenius-like temperature dependence (evidenced by the evolution of Hurst exponent shown in **Fig. 2**). Thus, the transition from unhindered avalanches glide at low temperatures to the fluctuation-pinned roughening phase at high temperatures manifests as the observed stress anomaly [38, 39]. More details about the avalanche dynamics are given in **Appendix A**.

## 3. Methodology

In this section, the theoretical framework of the fKPZ dynamics is established in **§3.1** ~ **§3.4**.

### 3.1. Kink solutions for disconnections

Rather than phenomenologically assuming the existence of discrete disconnections, we model the interface between two crystalline grains as a continuous 1D string parameterized by the spatial coordinate $x$ along the boundary. The local state of the interface is described by two continuous scalar fields, *i.e.*, the tangential shear displacement field $\phi(x,t)$ and the normal geometric height field $y(x,t)$.

Analogous to the intragranular slip, of which the energy landscape is defined as the standard generalized stacking fault energy (GSFE) as a function of the lattice vectors, we denote the GB energy landscape as the generalized GB energy (GGBE) surface, $\mathcal{V}(\phi,y)$, when two neighboring grains are displaced relative to each other. To capture the shear-migration coupling (where, the shear displacement necessitates normal migration to maintain the localized atomic registry) of the selected mode, the GGBE surface should incorporate cross-coupled periodic terms, thus can be described with Fourier harmonics,

$$\mathcal{V}(\phi,y) = \frac{V_0}{2}\left[1-\cos\left(\frac{2\pi\phi}{b_0}\right)\right] + \frac{U_0}{2}\left[1-\cos\left(\frac{2\pi y}{h_0}\right)\right] + W_0\left[1-\cos\left(\frac{2\pi\phi}{b_0}-\frac{2\pi y}{h_0}\right)\right] \tag{1}$$

where, $b_0$ and $h_0$ denote the characteristic lengths of the DSC lattice parallel and normal to the GB plane, respectively. $V_0$ and $U_0$ parameterize the energy barriers for pure shear and pure migration, respectively, while $W_0$ quantifies the strength of the crystallographic coupling constraint. The GGBE surface $\mathcal{V}(\phi,y)$ (units: $J/m^2$) thus describes the misfit energy between the two neighboring grains as a function of their relative displacement, as shown in **Fig. 4**a, with more details about string method [40] based evaluation in **Appendix B**.

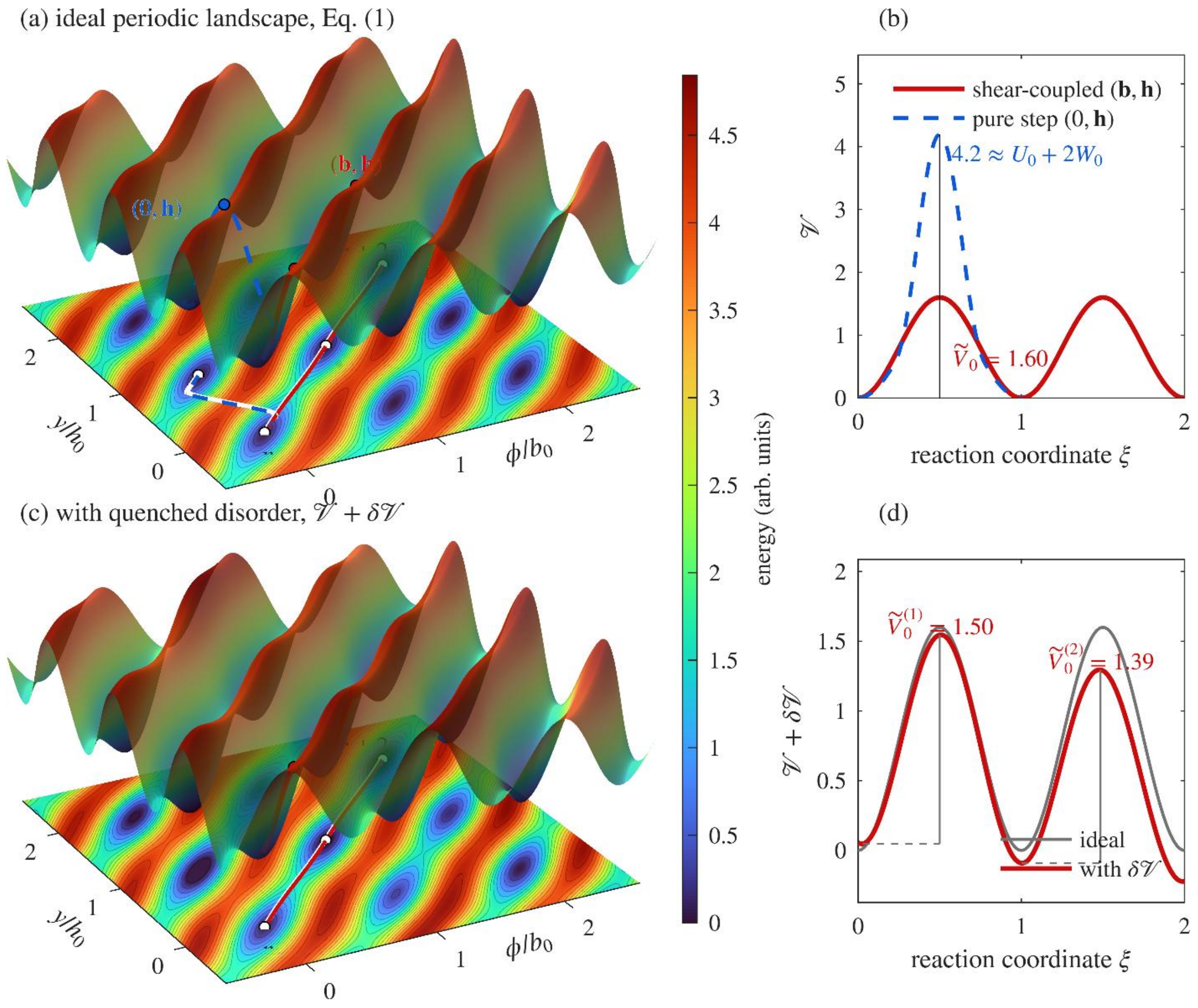

**Fig. 4**. Schematic GGBE surface of (a) ideally periodic lattice, and (c) with quenched disorder, respectively; and the MEP in (b) ideally GGBE landscape, and (d) with quenched disorder, respectively. The illustrative values used here are $V_0 = 1.0$, $U_0 = 0.6$ and $W_0 = 1.8$. It is noted that the MEP here is evaluated numerically using the string method [40], consistent with our theory **Eq**.(3).

The *Hamiltonian* of the system comprises elastic gradient terms and the above GGBE landscape $\mathcal{V}(\phi, y)$,

$$\mathcal{H}[\phi, y] = \int_{-\infty}^{+\infty} \left[\frac{1}{2}K_\phi(\partial_x\phi)^2 + \frac{1}{2}K_y(\partial_x y)^2 + \frac{1}{2}K_b(\partial_x^2 y)^2 + \mathcal{V}(\phi, y)\right] \mathrm{d}x \qquad (2)$$

where, $K_\phi$ is the effective shear stiffness (or interfacial stiffness, units: J/m$^2$), $K_y$ is the effective normal bending stiffness (capillary surface tension, units: J/m$^2$), and $K_b$ is the higher-order bending rigidity (units: J) originating from the discreteness of the crystal lattice.

It is noted that while the height field $y$ necessitates a higher-order bending rigidity $K_b(\partial_x^2 y)^2$ to account for the effect of local interfacial curvature, the shear field $\phi$ does not require a corresponding $(\partial_x^2 \phi)^2$ term. The shear field $\phi$ is regularized by the competition between the shear strain energy $K_\phi(\partial_x \phi)^2$ and the periodic potential $\mathcal{V}(\phi, y)$, restricting the tangential gradients to finite-width defect cores (solitons).

Since the GGBE surface $\mathcal{V}(\phi, y)$ follows the translational symmetry, *i.e.*, $\mathcal{V}(\phi + b_0, y + h_0) = \mathcal{V}(\phi, y)$, the shear-coupled migration requires that the shear displacement $\phi$ should be accompanied by the variation of normal height $y$ to maintain minimum energy state. By constraining the GB migration to the minimum energy path (MEP), *i.e.*, in the strong coupling limit ($W_0 \gg V_0, U_0$), the system is bound to the "valleys" of the GGBE landscape. The MEP thus satisfies the linear geometric constraint,

$$y(x) \approx \frac{h_0}{b_0} \phi(x) \tag{3}$$

According to the topological phase transition theory established by Chen, Srolovitz and Han (denoted as the ***CSH*** theory later) [15], this fixed relationship between the shear displacement ($\phi$) and normal migration ($y$) dominated by the single disconnection mode essentially retains the long-range elastic interactions below the Kosterlitz-Thouless (***KT***) transition temperature ($T_{KT}$), which is screened at high-temperature regimes. Substituting this constraint into the *Hamiltonian* reduces the 2D field theory to an effective 1D model. The local elastic energy collapses into the single shear field $\phi$ with an effective (shear) stiffness $K_{eff} = K_\phi + K_y(h_0/b_0)^2$. Besides, evaluating the GGBE along the MEP, yields $\mathcal{V}_{eff}(\phi) = \mathcal{V}(\phi)|_{MEP} = \frac{V_0+U_0}{2}[1 - \cos(2\pi\phi/b_0)]$. By defining the effective energy barrier as $\tilde{V}_0 = V_0 + U_0$, and neglecting the higher-order bending rigidity $K_b$ (which governs the short-range core width rather than dictating the overall profile of the defect), we obtain the *sine-Gordon-like Hamiltonian*,

$$\mathcal{H}_{eff}[\phi] = \int_{-\infty}^{+\infty} \left\{ \frac{1}{2} K_{eff} (\partial_x \phi)^2 + \frac{\tilde{V}_0}{2} \left[ 1 - \cos\left( \frac{2\pi\phi}{b_0} \right) \right] \right\} \mathrm{d}x \tag{4}$$

By applying the *principle of least action* ($\delta\mathcal{H}_{eff}/\delta\phi = 0$), we have the stationary states of the interface as,

$$K_{eff} \partial_{xx} \phi - \frac{\pi \tilde{V}_0}{b_0} \sin\left( \frac{2\pi\phi}{b_0} \right) = 0 \tag{5}$$

By applying the boundary conditions $\phi(-\infty) = 0$ and $\phi(+\infty) = b_0$, *i.e.*, the transition between adjacent minima of the GGBE potential, we can obtain the finite-energy kink solution,

$$\phi_{kink}(x) = \frac{2b_0}{\pi} \arctan\left[ \exp\left( \frac{x - x_0}{w} \right) \right] \tag{6}$$

where, $x_0$ is the soliton core position and $w = \frac{b_0}{\pi}\sqrt{\frac{K_{eff}}{2\tilde{V}_0}}$ is the characteristic core width. This demonstrates that the defect is not a singular point, but a (continuous) distortion field with finite width.

Evaluating the spatial integral of the gradient fields over the defect recovers the discrete crystallographic vectors. For the tangential displacement field $\phi(x)$,

$$\Delta\phi = \int_{-\infty}^{+\infty} \partial_x \phi_{kink} \mathrm{d}x = b_0 \equiv \boldsymbol{b} \tag{7}$$

$$\Delta y = \int_{-\infty}^{+\infty} \partial_x y \mathrm{d}x = \frac{h_0}{b_0}\Delta\phi = h_0 \equiv \boldsymbol{h} \tag{8}$$

where, the derivative of the kink profile is $\partial_x\phi_{kink} = \frac{b_0}{\pi w}\mathrm{sech}\left(\frac{x-x_0}{w}\right)$. The Burgers vector $\boldsymbol{b}$ and step height $\boldsymbol{h}$ that define a disconnection, are thus imposed by the DSC lattice. Besides, as shown in **Fig. 4**b, the development of a pure step [9] requires more energy than activating a disconnection.

### 3.2. Many-body *Langevin* equation

Having established that the elementary excitations of the interface are kinks with the crystallographic characteristic $(\boldsymbol{b}, \boldsymbol{h})$, we now move to the mesoscale regime. As shown in **Fig. 5**, the GB is abstracted as a 2D domain (along the sliding direction $x$ and the transverse coordinate $l$) containing $N$ disconnections, with the instantaneous position of the $i$-th disconnection parameterized as $x_i(l,t)$.

#### *3.2.1. Generalized Peach-Koehler-like force via the Principle of Virtual Work*

Considering a unit-length segment of an individual disconnection subjected to the applied shear stress $\tau$ (units: Pa) and thermodynamic driving force $P$ (the free energy difference per unit volume across the GB, units: J/m$^3$), and letting the disconnection undergoes a virtual displacement $\delta x$, we can have,

*1) Mechanical virtual work*. The disconnection glide causes a tangential displacement of magnitude $b$ between the upper and lower grains, with the sheared area is $1 \cdot \delta x$. The mechanical work performed by the applied shear stress $\tau$ is thus,

$$\delta W_{mech} = (\tau \cdot 1 \cdot \delta x) \cdot b = \tau b \delta x \tag{9}$$

*2) Thermodynamic virtual work*. The virtual displacement $\delta x$ causes the step of height $h$ to sweep a volume $\delta V = h \cdot 1 \cdot \delta x$. The material within this volume transforms from the higher-energy grain to the lower-energy one, releasing the free energy,

$$\delta W_{thermo} = P\delta V = Ph\delta x \tag{10}$$

According to the *Principle of Virtual Work*, the total virtual work should be equal to the work done by an effective driving force $F_{ext}$, $F_{ext}\delta x = \delta W_{mech} + \delta W_{thermo}$. We thus have,

$$F_{ext} = \tau b + Ph \tag{11}$$

**Eq.**(11) describes how the thermodynamic pressure $P$ (perpendicular to the GB) drives the defect motion horizontally, *i.e.*, the step height $h$ acts as a *geometric converter*, transforming the normal driving into the tangential thrust through the kinematic coupling defined in the GGBE landscape.

### *3.2.2. Nonlocal interactions and overdamped Langevin dynamics*

In crystals, the disconnection motion is damped by phonon scattering and atomic reorganization, rendering the inertial term negligible. We thus model the motion of the $i$-th disconnection (see **Fig. 5**) via the overdamped *Langevin* equation [41],

$$\beta \dot{x}_i = (\tau b_i + Ph_i) + F_{int}(x_i) + F_{pin}(x_i) + F_{self}(x_i) + \xi_i(l,t) \tag{12}$$

where, $\beta$ is the *phonon drag coefficient* (usually on the order of ~ $10^{-5}$ Pa·s for metals [42]), quantifying the viscous resistance to the disconnection motion, and $l$ is the coordinate along the disconnection line.

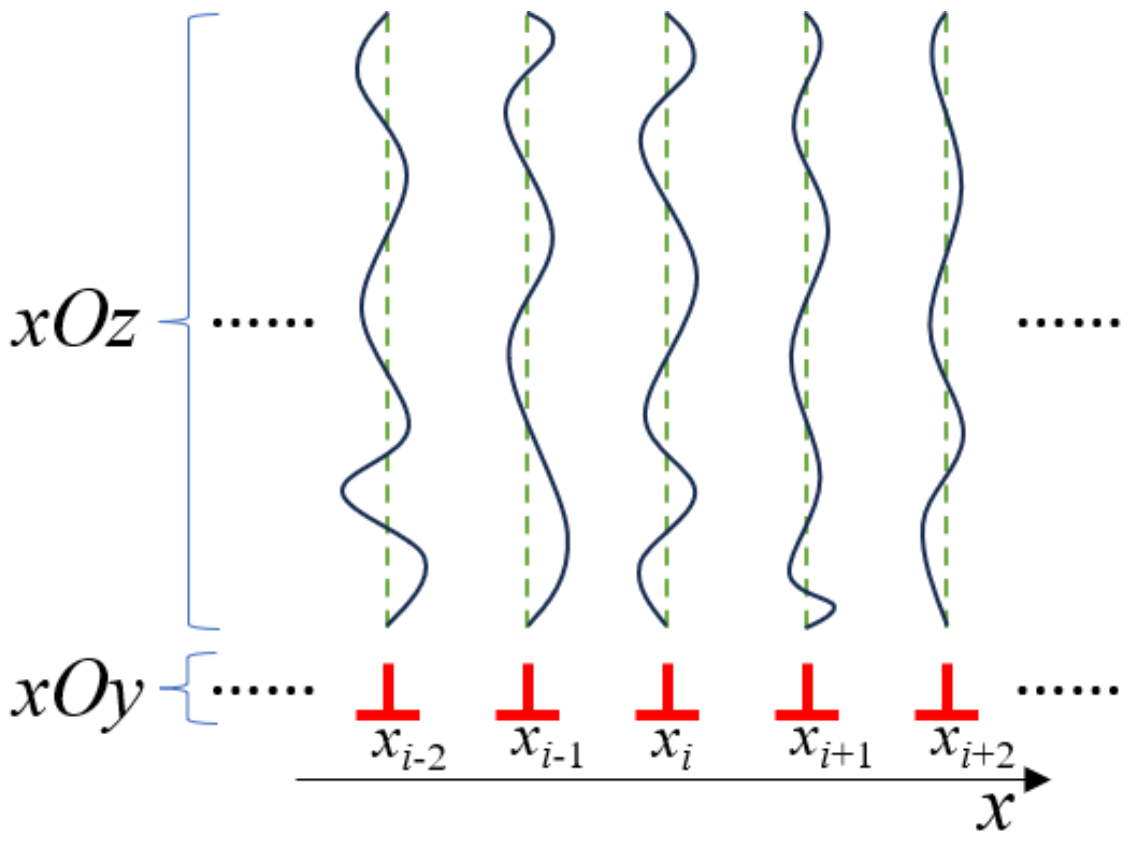

**Fig. 5**. Schematics of the disconnection ensemble considered in **Eq.**(12).

While the first term in the r.h.s. of **Eq.**(12) is the external loading, the other four terms are,

*1) Elastic interactions*. As established in the dislocation theory, parallel defects with the Burgers vector $b$ repel each other via the stress field that decays as $1/r$. Invoking the small-gradient approximation ($|\partial_l x_i| \ll 1$) for the defect profile, the interaction force on defect $i$ due to all other defects $j$, is given by the elastic kernel averaged over the characteristic length $L_z$,

$$F_{int}[x_i(l,t)] = \frac{K}{L_z} \sum_{j \neq i} \int \frac{b_i b_j}{x_i(l,t) - x_j(l',t)} dl' \tag{13}$$

where, $K \equiv G/[2\pi(1-\nu)]$, with $G$ is the shear modulus and $\nu$ is Poisson's ratio. The prefactor $1/L_z$ ensures the dimensional consistency of a force per unit length (N/m).

*2) Localized periodic pinning*. While the continuous field theory in **§3.1** yields translationally invariant solitons, the discrete nature of the crystal lattice breaks this continuous symmetry. The energy of a defect centered at $x_i$ should be evaluated as a sum over the discrete lattice rather than a continuous integral, *i.e.*, $E_{disc}(x_i) = \sum_n a \cdot \mathcal{H}_{eff}[\phi_{kink}(na - x_i)]$, where $a$ is the periodicity of the DSC lattice along the GB. Applying the Poisson summation formula, this discrete summation decouples into the translationally invariant energy $E_{cont}$ and a spatially periodic residual, which constitutes the *Peierls-like* pinning potential $V_P(x_i)$,

$$V_P(x_i) \approx \frac{\Delta E_P}{2}\left[1 - \cos\left(\frac{2\pi x_i}{a}\right)\right] \tag{14a}$$

For the *sine-Gordon-like* GGBE derived in **§3.1**, the amplitude of the pinning barrier scales exponentially with the core-width $w$ in **Eq.**(6), *i.e.*, $\Delta E_P \propto \tilde{V}_0 \exp(-\pi^2 w/a)$. Thus, the discrete lattice sampling of the GGBE gives the localized restoring force per unit length as,

$$F_{pin}(x_i) = -\frac{\partial V_P(x_i)}{\partial x_i} \approx -\frac{\pi \Delta E_P}{a} \sin\left(\frac{2\pi x_i}{a}\right) \tag{14b}$$

**Eq.**(14b) reveals that the periodic pinning in the *Langevin* dynamics originates from the discrete sampling of the continuum GGBE potential.

*3) Defect line tension and self-energy*. When the disconnection line undergoes local bending distortions along the coordinate $l$, its (intrinsic) core energy and line tension generate the restoring force,

$$F_{self}(x_i) = K_y^{\text{mic}} \frac{\partial^2 x_i}{\partial l^2} - K_b^{\text{mic}} \frac{\partial^4 x_i}{\partial l^4} \tag{15}$$

where, $K_y^{mic}$ and $K_b^{mic}$ are the capillary tension and (higher-order) bending rigidity parameters of the disconnection line, respectively.

*4) Thermal noise*. The stochastic force (per unit length) of the disconnection line $\xi_i(l,t)$ originates from the thermal fluctuation induced by the phonon bath. According to the FDT of non-equilibrium statistical mechanics, this thermal noise should be Gaussian with zero mean and no spatiotemporal correlations,

$$\langle \xi_i(l,t) \rangle = 0 \tag{16a}$$

$$\langle \xi_i(l,t)\, \xi_j(l',t') \rangle = 2\beta k_B T\, \delta_{ij} \delta(l-l')\delta(t-t') \tag{16b}$$

where, $k_B$ is Boltzmann's constant, $T$ is the absolute temperature, the spatial delta function $\delta(l-l')$ (units: 1/m) and temporal delta function $\delta(t-t')$ (units: 1/s) render the noise correlation dimensionally consistent with the square of the force per unit length, (N/m)$^2$.

## 3.3. Coarse-graining to the fKPZ equation

In this section, we construct the SPDE governing the morphological evolution of the GB profile $y(x,t)$, starting from the many-body *Langevin* dynamics of disconnections established in **§3.2**. To connect the discrete defect positions and the continuous height field, we first establish the kinematic constraints (**§3.3.1**), then evaluate the spatial coarse-graining of the nonlocal elasticity (**§3.3.2**), finally derive the migration velocity and the KPZ nonlinearity from the external loading (**§3.3.3**) and the local restoring forces (**§3.3.4**).

### *3.3.1. Disconnection flux along the GB*

By assigning the disconnections with topological charges $\sigma_i = \pm 1$ (where, $b_i = \sigma_i b_0$), we can define the net density field $\rho(x,t)$ (units: m$^{-1}$) and the net flux field $\mathcal{J}(x,t)$ (units: s$^{-1}$) of disconnections along the *x*-axis in the laboratory reference frame,

$$\rho(x,t) = \textstyle\sum_{i=1}^{N} \sigma_i \delta(x - \bar{x}_i(t)) \tag{17}$$

$$\mathcal{J}(x,t) = \textstyle\sum_{i=1}^{N} \sigma_i \dot{\bar{x}}_i(t) \delta(x - \bar{x}_i(t)) \tag{18}$$

where, $\bar{x}_i(t) = \int_0^{L_z} x_i(l,t) dl / L_z$ is the line-averaged position of the $i$-th disconnection, neglecting the fluctuations over the characteristic length $L_z$. The glide velocity $\dot{\bar{x}}_i(t)$ is described by the overdamped *Langevin dynamics* established in **Eq**.(12). Since the topological charge $\sigma_i$ of the disconnection ensemble is conserved, the evolution of the density field $\rho(x,t)$ satisfies the continuity equation,

$$\frac{\partial \rho(x,t)}{\partial t} + \frac{\partial \mathcal{J}(x,t)}{\partial x} = 0 \tag{19}$$

While the accumulation of disconnections contributes to the spatial gradient of the GB profile $y(x,t)$, a positive disconnection forms a downward step along the positive *x*-axis, giving, $\frac{\partial y(x,t)}{\partial x} = -h_0\rho(x,t)$ according to the Volterra construction [43]. By differentiating this equation with respect to time $t$ and substituting into **Eq.**(19), we can link the normal growth rate of the interface to the lateral flux,

$$\frac{\partial y(x,t)}{\partial t} = h_0\mathcal{J}(x,t) \tag{20}$$

Revisiting the *Langevin dynamics* **Eq.**(12), the local glide velocity of each disconnection is thus driven by the total force, *i.e.*, $\dot{\bar{x}}_i = F_{total,i}/\beta$. Substituting this relation into **Eq.**(18), the net flux is governed by the superposition of all force components acting on the defect ensemble,

$$\mathcal{J}(x,t) = \frac{1}{\beta}\sum_{i=1}^{N}\sigma_i F_{total,i}\delta(x-\bar{x}_i(t)) \tag{21a}$$

To homogenize this discrete summation into a continuum field, we would evaluate the sign dependence of the localized forces. In dislocation mechanics, the driving and restoring forces acting on a defect (e.g., *Peach-Koehler-like* forces) are proportional to its topological charge. Therefore, the total force on the *i*-th disconnection can be recast as $F_{total,i} = \sigma_i F_{total}(\bar{x}_i)$, where $F_{total}(x,t)$ represents the homogenized force acting on a unit defect with a positive sign ($\sigma = +1$). Substituting this relation into **Eq.**(21a) yields,

$$\mathcal{J}(x,t) = \frac{F_{total}(x,t)}{\beta}\sum_{i=1}^{N}\sigma_i^2\delta(x-\bar{x}_i(t)) \tag{21b}$$

Since the topological charge is restricted to $\sigma_i = \pm 1$, its square reduces to $\sigma_i^2 \equiv 1$, guaranteeing that oppositely-signed defects, despite gliding in opposite directions, contribute to the GB migration additively.

In the continuum limit, the discrete sum $\sum\delta(x-\bar{x}_i)$ constitutes the total defect density field, denoted as $\rho_{tot}(x,t)$ to distinguish it from the net density $\rho(x,t)$. By defining the Eulerian velocity field as $v(x,t) = F_{total}(x,t)/\beta$, the flux simplifies to,

$$\mathcal{J}(x,t) = \rho_{tot}(x,t)v(x,t) \tag{22}$$

Substituting this flux into **Eq.**(20), we obtain the master equation for the GB migration,

$$\frac{\partial y(x,t)}{\partial t} = h_0\rho_{tot}(x,t)v(x,t) \tag{23}$$

where, the velocity field $v(x,t)$ is governed by the superposition of all coarse-grained force components,

$$v(x,t) = [F_{drive} + F_{int}(x,t) + F_{pin}(x,t) + F_{self}(x,t) + \xi(x,t)]/\beta \quad (24)$$

To construct the final SPDE, we would couple each term of $v(x,t)$ with the interfacial kinematics. The theoretical roadmap is organized as follows,

**§3.3.2** discusses the spatial coarse-graining of the nonlocal elasticity $F_{int}$.

**§3.3.3** formulates the interplay between the external loading $F_{drive}$ and the interfacial kinematics, deriving the migration velocity and the KPZ nonlinearity. It is noted that the coupling between the interface slope and GB migration is resolved via the arc-length formulation (yielding the KPZ nonlinearity) here.

**§3.3.4** derives the restoring forces ($F_{pin}$ and $F_{self}$) by solving the functional derivative of the *Hamiltonian*.

*3.3.2. From the $1/r$ interaction to the fractional Laplacian*

This subsection focuses on the evaluation of the nonlocal elastic field $F_{int}(x,t)$ in **Eq.**(24). In the long-wavelength limit ($q \to 0$), the line-averaged position $\bar{x}_i(t)$ established in **§3.3.1** serves as the baseline for evaluating the line integral in **Eq.**(13). We decompose the local coordinate of any individual disconnection string into its mean position and a zero-mean fluctuation along the out-of-plane *z*-axis, *i.e.*, $x_i(l,t) = \bar{x}_i(t) + \delta x_i(l,t)$.

By assuming that the inter-disconnection distance in the *xOy* plane ($\Delta\bar{x}_{ij} = \bar{x}_i - \bar{x}_j$) is much larger than the amplitude of the out-of-plane fluctuation, the leading-order term representing the geometric fluctuations integrates to zero over the characteristic length $L_z$ through the perturbative expansion of the integrand. Thus, the nonlocal interaction force per unit length within the *xOy* plane reads as,

$$F_{int,i} \approx \sum_{j \neq i} \frac{K b_i b_j}{\bar{x}_i(t) - \bar{x}_j(t)} \quad (25a)$$

As established in **§3.3.1**, the localized force should be the product of the topological charge and continuum field, *i.e.*, $F_{int,i} = \sigma_i F_{int}(\bar{x}_i, t)$. By decomposing the Burgers vectors via their topological charges, $b_i b_j = \sigma_i \sigma_j b_0^2$, we can factor out the topological charge of the subjected defect ($\sigma_i$) from the summation,

$$F_{int,i} = \sigma_i \left[ K b_0^2 \sum_{j \neq i} \frac{\sigma_j}{\bar{x}_i(t) - \bar{x}_j(t)} \right] \quad (25b)$$

where, the bracketed term is the discrete form of the elastic field acting on a unit defect with positive sign ($\sigma = +1$) at location $\bar{x}_i$. The remaining topological charges ($\sigma_j$) are subsumed into the net disconnection density field $\rho(x',t)$ defined in **Eq.**(17). Thus, the summation reduces to a spatial convolution integral, *i.e.*, the nonlocal elastic interaction force, $F_{int}(x,t)$, derived as the Cauchy principal value integral,

$$F_{int}(x,t) = Kb_0^2 \int_{-\infty}^{+\infty} \frac{\rho(x',t)}{x-x'} \mathrm{d}x' \tag{26}$$

Considering the kinematic relation between the net defect density and the interface slope, we have $\rho(x,t) = -\partial_x y(x,t)/h_0$, where the negative slope reflects the accumulation of positive disconnections. Substituting this relation into **Eq.**(26) yields,

$$F_{int}(x,t) = -\frac{Kb_0^2}{h_0} \int_{-\infty}^{+\infty} \frac{\partial_{x'} y(x',t)}{x-x'} \mathrm{d}x' \tag{27}$$

To solve this singular integral, we employ the spatial Fourier transform (from real-space coordinate $x$ to wavevector $q$) by applying the convolution theorem $\mathcal{F}[f * g] = \mathcal{F}[f] \cdot \mathcal{F}[g]$, *i.e.*,

a) the Fourier transform of $1/x$ is $\mathcal{F}[1/x] = -i\pi \operatorname{sgn}(q)$, where, $\operatorname{sgn}(q)$ is the sign function.

b) the Fourier transform of the spatial derivative is $\mathcal{F}[\partial_x y] = iq\tilde{y}(q)$.

Multiplying these components in the wavevector domain yields,

$$\mathcal{F}[F_{int}] = -\frac{Kb_0^2}{h_0}(-i\pi \operatorname{sgn}(q)) \cdot (iq\tilde{y}(q)) = -\frac{\pi K b_0^2}{h_0} |q|\, \tilde{y}(q) \tag{28}$$

where, the multiplier $|q|$ is the spectral representation of the *half-Laplacian*, $(-\Delta)^{1/2}$. By inverting the Fourier transform back to real space, we have the nonlocal force distribution as,

$$F_{int}(x,t) = -\frac{\pi K b_0^2}{h_0} (-\Delta)^{1/2}\, y(x,t) \tag{29}$$

According to the *Orowan-like* kinematic relation, the interface velocity induced by the nonlocal elasticity is defined as $\partial_t y|_{int} = h_0 \rho_{tot} v_{int}$, where $v_{int} = F_{int}/\beta$ is the local glide velocity.

To a first approximation, we decompose the total defect density into the homogeneous background ($\rho_0$) and the local fluctuation ($\delta\rho_{tot}$), *i.e.*, $\rho_{tot}(x,t) = \rho_0 + \delta\rho_{tot}(x,t)$. For a nominally-flat interface, while the net density vanishes ($\rho(x,t) = 0$), the background density $\rho_0$ remains positive due to the presence of thermal or structural disconnection dipoles. Substituting this decomposition into the kinematic relation, coupled with the nonlocal elasticity, yields a higher-order nonlinear term proportional to $\delta\rho_{tot}(-\Delta)^{1/2} y$.

However, in the long-wavelength limit, the leading-order KPZ nonlinearity ($\propto (\partial_x y)^2$), arising from the geometric projection of the external loading (see **§3.3.3**), dominates the morphological evolution in the post-crossover regime. Thus, the higher-order elastic fluctuations become subdominant, allowing us to truncate the elastic interactions by replacing $\rho_{tot}(x,t)$ with the homogeneous background $\rho_0$. Substituting the nonlocal force $F_{int}$ into this relation allows the step height $h_0$ to cancel out naturally, yielding,

$$\left.\frac{\partial y}{\partial t}\right|_{int} = -\frac{\rho_0 \pi K b_0^2}{\beta}(-\Delta)^{1/2}\, y(x,t) \equiv -\mu_{el}(-\Delta)^{1/2}\, y(x,t) \tag{30}$$

where, $\mu_{el} = \rho_0 G b_0^2/(2\beta(1-v))$ is the effective nonlocal elastic mobility. **Eq.**(30) establishes that the $1/r$ elastic interaction between disconnections emerges as the *nonlocal (fractional) diffusion operator* [44] acting on the interface height field.

*3.3.3. The KPZ nonlinear coefficient λ*

Next, we formulate the contribution of the external driving force. According to the relations established in **§3.3.1** ($b_i = \sigma_i b_0$, $h_i = \sigma_i h_0$), the driving force acting on an elementary disconnection is given by $F_{drive} = \tau b_0 + P h_0$. We now derive the KPZ nonlinear coupling coefficient $\lambda$, by substituting the *Orowan-like* kinematics into the small-slope expansion of the interface, *i.e.*,

*Step 1: Flat-interface reference velocity*. To establish the reference sliding kinetics, we consider a nominally-flat interface ($\partial_x \mathrm{y} = 0$) moving at a reference velocity $v_0$. It is noted that, $v_0$ is the effective migration velocity determined by the interplay between the driving force $F_{drive}$ and the lattice pinning. In the pre-crossover state, the interface is structurally arrested, *i.e.*, $v_0 \cong 0$. However, upon the crossover, the interface escapes from the periodic landscape, recovering the steady-state migration $v_0 > 0$.

*Step 2: Nonlinear kinematics of the tilted interface*. As shown in **Fig. 6**, when the interface exhibits a local tilt (characterized by the slope $m = \partial_x y \neq 0$), the glide plane deviates by an angle $\theta = \arctan(\partial_x y)$ from the reference state. In this "necklace" model, the horizontal component of the disconnection glide velocity is fixed by the applied shear loading, while the local inclination affects the glide velocity and the projected defect density as follows,

(i) *Kinematic velocity amplification*. As shown in the "necklace" model schematized in **Fig. 6**, the discrete disconnections are constrained to glide along the locally inclined tangent. To accommodate the horizontal shear imposed globally, the horizontal projection of the disconnection gliding velocity ($v_{glide}$) should match the steady-state sliding velocity ($v_{slide,x}$, *i.e.*, the sliding velocity of a disconnection core along the *x*-axis direction), *i.e.*, $v_{slide,x} = v_{glide}\cos\theta$. This causes the disconnection gliding velocity ($v_{glide}$) to amplify nonlinearly on tilted segments, $v_{glide} = v_{slide,x}/\cos\theta = v_{slide,x}\sqrt{1+(\partial_x y)^2}$ (see **Fig. 6**b).

(ii) *Derivation of the growth rate*. Considering the conservation of the local defect number $N$, the density $\rho_{tot}(x,t) \equiv \mathrm{d}N/\mathrm{d}x$ should be evaluated from the line density $\mathrm{d}N/\mathrm{d}s$ defined along the interfacial arc-length $s$. Since the crystallographic spacing between disconnections remains invariant regardless of the local tilt (in the "necklace" model shown in **Fig. 6**), the line density is equal to the background density $\rho_0$ of the flat reference state (*i.e.*, $\mathrm{d}N/\mathrm{d}s \equiv \rho_0$). By applying the differential relation $\mathrm{d}s = \sqrt{1+(\partial_x y)^2}\mathrm{d}x$, the density field is derived through the chain rule,

$$\rho_{tot}(x,t) = \frac{\mathrm{d}N}{\mathrm{d}s}\frac{\mathrm{d}s}{\mathrm{d}x} = \rho_0\sqrt{1+(\partial_x y)^2} \tag{31a}$$

Substituting the above **Eq**.(31a) and the local velocity $v(x,t) \equiv v_{glide}$ into **Eq**.(23), yields,

$$\frac{\partial y}{\partial t} = h_0\left[\rho_0\sqrt{1+(\partial_x y)^2}\right]v_{glide} \tag{31b}$$

As established in **§3.3.1**, the local migration velocity $v_n$, normal to the locally inclined interface (see **Fig. 6**c), is determined by the locally amplified defect glide flux, *i.e.*, $v_n \equiv h_0\rho_0 v_{glide}$. The vertical growth rate of the interface height profile in the laboratory reference frame is thus given by,

$$\frac{\partial y}{\partial t} = v_n\sqrt{1+(\partial_x y)^2} \tag{31c}$$

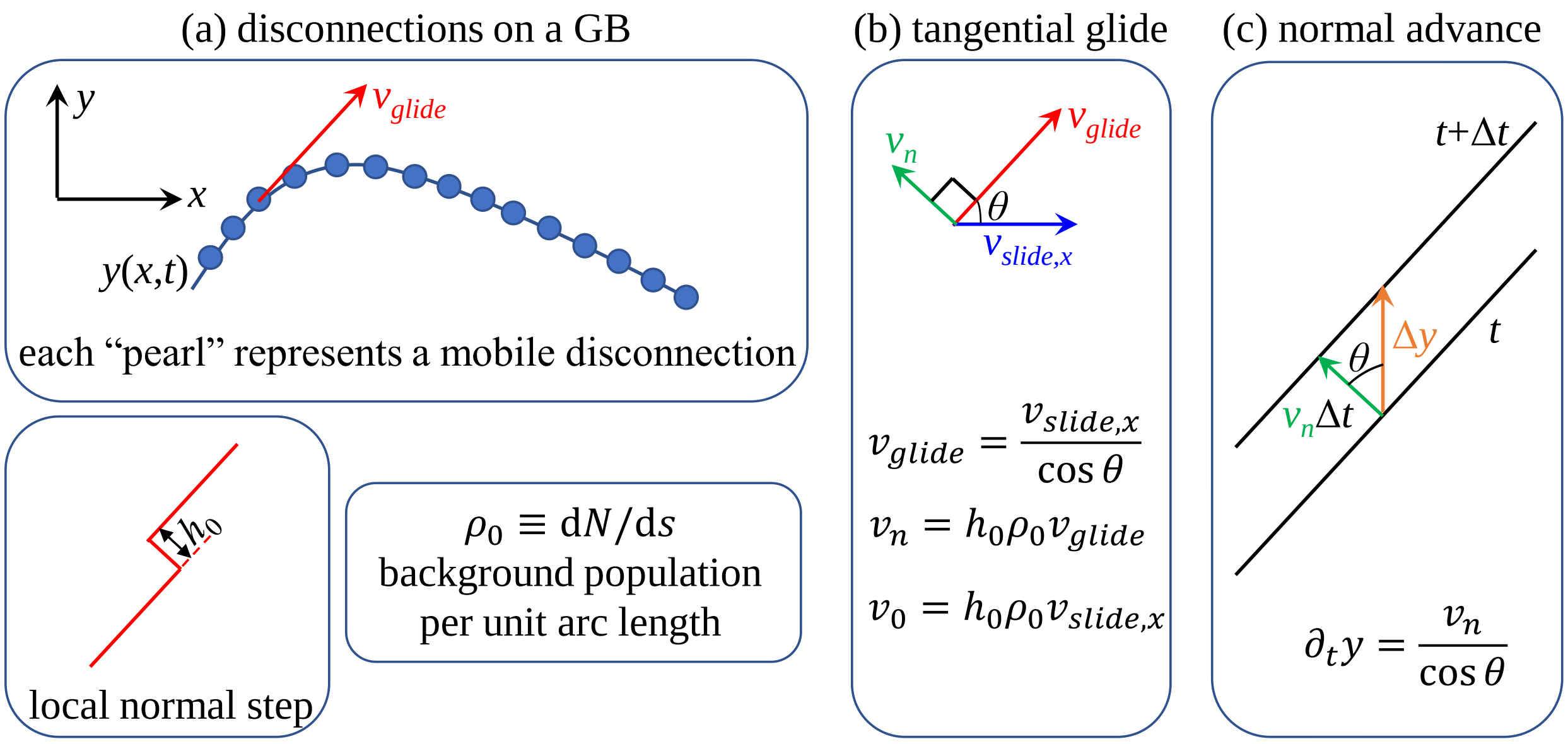

**Fig. 6**. Schematics of the disconnection kinematics.

*Step 3: Homogenization of the velocity field*. To solve the velocity field, we define the reference velocity of the nominally-flat interface driven by the steady-state *Peach-Koehler-like* force as, $v_0 = h_0\rho_0 v_{slide,x}$, where $\rho_0$ is the homogeneous background density defined in **§3.3.2**. When the mean pinning contribution is negligible in the steady-sliding regime, this reference velocity approaches $v_0 \cong v_0^e = \frac{h_0\rho_0}{\beta}(\tau b + Ph)$, where the superscript '$e$' indicates external loading. As established by the kinematic compatibility condition in *Step 2*, the local normal velocity $v_n = h_0\rho_0 v_{glide}$ on the tilted segment containing the geometric factor $1/\cos\theta$ from the amplified defect glide,

$$v_n = h_0\rho_0\left(\frac{v_{slide,x}}{\cos\theta}\right) = \frac{v_0}{\cos\theta} = v_0\sqrt{1+(\partial_x y)^2} \tag{32a}$$

By expanding this multiplier in the small-slope regime ($|\partial_x y| \ll 1$) up to the second order, we have the local migration velocity as,

$$v_n \cong v_0\left[1+\frac{1}{2}(\partial_x y)^2 + \mathcal{O}((\partial_x y)^4)\right] \tag{32b}$$

Substituting **Eq.**(32b) into **Eq.**(31c) and expanding to $\mathcal{O}((\nabla y)^2)$,

$$\begin{aligned}\frac{\partial y}{\partial t} &\cong v_0\left[1+\frac{1}{2}(\partial_x y)^2\right]\sqrt{1+(\partial_x y)^2} \\ &= v_0\left[1+\frac{1}{2}(\partial_x y)^2\right]\left[1+\frac{1}{2}(\partial_x y)^2+\cdots\right] \\ &= v_0 + v_0(\partial_x y)^2 + \mathcal{O}((\partial_x y)^4)\end{aligned} \tag{33}$$

*Step 4: Identification of the KPZ nonlinear coefficient*. By comparing the small-slope expansion **Eq.**(33) with the kinematic form of the KPZ equation in the laboratory reference frame, $\partial_t y = v_0 + \frac{\lambda}{2}(\partial_x y)^2 + \cdots$ [22, 45], we can identify the KPZ coupling coefficient as, $\lambda = 2v_0$. This relation provides the microscopic expression for the KPZ nonlinearity, *i.e.*, $\lambda$ is proportional to the (steady-state) interface migration velocity $v_0$, meaning it is activated only in the post-crossover regime ($v_0 > 0$) and vanishes when the interface is pinned in the pre-crossover regime ($v_0 \cong 0$).

### *3.3.4. Variational derivation of restoring forces*

To complete the evaluation of $F_{total}$, we further derive the localized restoring forces, namely, the periodic pinning $F_{pin}$ and the self-energy $F_{self}$ from the variation of the free-energy functional. The *Hamiltonian* characterizing the local energy landscape of the GB height profile $y(x,t)$ is defined as,

$$\mathcal{H}_{local}[y] = \int\left[\mathcal{V}_{sub}(y) + \frac{1}{2}K_y\left(\frac{\partial y}{\partial x}\right)^2 + \frac{1}{2}K_b\left(\frac{\partial^2 y}{\partial x^2}\right)^2\right]\mathrm{d}x \tag{34}$$

where, the substrate potential $\mathcal{V}_{sub}(y) = \frac{\tilde{V}_0}{2}\left[1 - \cos\left(\frac{2\pi y}{h_0}\right)\right]$ is formally identical to the effective shear potential $\mathcal{V}_{eff}(\phi)$ established in **§3.1**. By applying the MEP constraint $y = (h_0/b_0)\phi$, the coupled GGBE landscape is re-parameterized into the normal migration coordinate, yielding this substrate potential.

The thermodynamic restoring pressure $P_{res}$ (force per unit area, units: N/m$^2$) acting on the GB is defined by the variational derivative, $P_{res}(x,t) = -\delta\mathcal{H}_{local}/\delta y$. Through the *Principle of Virtual Work* established in **§3.2.1**, we have the restoring force (per unit length) on an individual disconnection string as $F_{res,i} = P_{res}h_0 = -h_0(\delta H_{local}/\delta y)$. According to the *Orowan*-like kinematics under the weak density fluctuation approximation ($\rho_{tot} \approx \rho_0$), the GB velocity contributed by these restoring forces is $\partial_t y|_{local} = h_0\rho_0 v_{res}$, where the corresponding glide velocity is $v_{res} = F_{res,i}/\beta$. By substituting the restoring force into this *Orowan*-like equation, we can obtain the gradient flow as,

$$\left.\frac{\partial y}{\partial t}\right|_{\text{local}} = \frac{h_0\rho_0}{\beta}\left(-h_0\frac{\delta\mathcal{H}_{local}}{\delta y}\right) \equiv -M\frac{\delta\mathcal{H}_{\text{local}}}{\delta y} \tag{35}$$

where, $M \equiv \rho_0 h_0^2/\beta$ is the interfacial mobility. The quadratic dependence on the step height $h_0^2$ embodies its dual roles as, *i.e.*, 1) converting the interfacial pressure to lateral defect force, and 2) projecting the defect flux into interfacial growth. By evaluating the functional variations for each constituent of $\mathcal{H}_{local}$, we can obtain three field operators as follows,

*1) Periodic lattice pinning*. The functional derivative of the substrate potential gives the periodic restoring velocity, which drives the interface toward the CSL energy minima,

$$\left.\frac{\partial y}{\partial t}\right|_{pin} = -M\frac{\delta}{\delta y}\int \mathcal{V}_{sub}(y)dx = -v_p\sin\left(\frac{2\pi y}{h_0}\right) \tag{36}$$

where, $v_p = \pi M\tilde{V}_0/h_0$ is the characteristic depinning velocity.

*2) Capillary interfacial tension* ($\nu_2$). The functional derivative of the gradient energy term (with the interfacial stiffness $K_y$ obtained from defect line tension $K_y^{mic}$) gives the second-order *Laplacian* diffusion,

$$\left.\frac{\partial y}{\partial t}\right|_{cap} = -M\frac{\delta}{\delta y}\int\frac{1}{2}K_y(\partial_x y)^2dx = MK_y\frac{\partial^2 y}{\partial x^2} \equiv \nu_2\frac{\partial^2 y}{\partial x^2} \tag{37}$$

*3) Hyperviscous damping* ($\nu_4$). The functional derivative of the higher-order term (homogenized from the bending rigidity $K_b^{mic}$ of 1D disconnection line to the curvature modulus $K_b$ of the 2D interface) provides the fourth-order hyperviscous dissipation to suppress the sub-atomic fluctuations,

$$\left.\frac{\partial y}{\partial t}\right|_{hyper} = -M\frac{\delta}{\delta y}\int \frac{1}{2}K_b(\partial_x^2 y)^2 dx = -MK_b\frac{\partial^4 y}{\partial x^4} \equiv -\nu_4\frac{\partial^4 y}{\partial x^4} \tag{38}$$

*3.3.5. The complete fKPZ equation*

By assembling all components derived above, *i.e.*, the nonlocal elasticity $-\mu_{el}(-\Delta)^{1/2}$ defined in **§3.3.2**, the steady-state velocity $v_0$ and the KPZ nonlinearity $\frac{\lambda}{2}(\partial_x y)^2$ in **§3.3.3**, and the variational forces in **§3.3.4**, we can obtain the *fKPZ* equation,

$$\frac{\partial y}{\partial t} = v_0^e - \mu_{el}(-\Delta)^{1/2}y - v_p \sin\left(\frac{2\pi y}{h_0}\right) + \nu_2\frac{\partial^2 y}{\partial x^2} - \nu_4\frac{\partial^4 y}{\partial x^4} + \frac{\lambda}{2}\left(\frac{\partial y}{\partial x}\right)^2 + \eta_{tot}(x,t,y) \tag{39}$$

It is noted that Eq.(39) should be restricted to the low-temperature single disconnection mode regimes (constrained by Eq.(3)) below the ***KT*** transition temperature established in the ***CSH*** theory, above which multiple modes are activated and the long-range elasticity would be screened [15]. The total noise field $\eta_{tot}$ acting on the interface can be decomposed as two parts, *i.e.*, the thermal fluctuations $\eta_{th}(x,t)$ and the (static) quenched disorder embedded within the lattice $\eta_q\big(x,y(x,t)\big)$, *i.e.*, $\eta_{tot}(x,t,y) = \eta_{th}(x,t) + \eta_q\big(x,y(x,t)\big)$, which are detailed as follows,

*1) Gaussian white noise* ($\eta_{th}$). Inspired by the stochastic dynamics of dislocations [46], it is assumed that the thermal-noise component describes fluctuations induced by the phonon bath. By defining the random force density under the continuum sense (*i.e.*, $\delta_{ij} \rightarrow \delta(x-x')/\rho_0$), the FDT established in **Eq.**(16b) gives the microscopic noise correlation as $\langle\xi(x,t)\xi(x',t')\rangle = \frac{2\beta k_B T}{\rho_0 L_z}\delta(x-x')\delta(t-t')$ over the characteristic length $L_z$. Then, this random force is projected onto the normal migration velocity via the established relation, *i.e.*, $\eta_{th}(x,t) = h_0\rho_0 v_\xi = \frac{h_0\rho_0}{\beta}\xi(x,t)$. By evaluating the macroscopic noise correlation, we have,

$$\langle\eta_{th}(x,t)\eta_{th}(x',t')\rangle = \left(\frac{h_0\rho_0}{\beta}\right)^2\frac{2\beta k_B T}{\rho_0 L_z}\delta(x-x')\delta(t-t') \tag{40a}$$

By using the interfacial mobility defined in **§3.3.4** ($M \equiv \rho_0 h_0^2/\beta$), **Eq.**(40a) reduces to,

$$\langle\eta_{th}(x,t)\eta_{th}(x',t')\rangle = \frac{2Mk_B T}{L_z}\delta(x-x')\delta(t-t') \tag{40b}$$

*2) Quenched disorder* ($\eta_q$) *and motional averaging* ($\eta_{dis}$). As observed in MD simulations, in the pre-crossover regime ($v_0 \cong 0$), the interface is pinned by the periodic lattice potential and disconnection pile-ups. The thermal noise becomes asymptotically irrelevant under the renormalization group (RG) framework, leaving the interfacial dynamics dominated by the quenched disorder $\eta_q(x,y)$. The spatial statistics of this quenched landscape are characterized by the short-range correlations along the migration ($y$-axis) direction, coupled with the long-range correlation within the interface,

$$\langle \eta_q(x,y)\eta_q(x',y')\rangle = \Delta_q(|x-x'|)R_{\ell_y}(y-y') \tag{40c}$$

where, $R_{\ell_y}$ is the normalized correlation kernel with the finite correlation length $\ell_y$ along the migration direction, while $\Delta_q(|x-x'|)$ captures the (in-plane) spatial correlation of the quenched heterogeneities along the interface. To a first approximation, we adopt the Gaussian regularization as follows [47],

$$R_{\ell_y}(s) = \frac{1}{\sqrt{2\pi}\ell_y}\exp\left(-\frac{s^2}{2\ell_y^2}\right), \quad \int_{-\infty}^{\infty} R_{\ell_y}(s)\mathrm{d}s = 1 \tag{40d}$$

The limit $R_{\ell_y}(s) \to \delta(s)$ as $\ell_y \to 0^+$ holds in the distributional sense, while the delta-correlated form is used only for the subsequent coarse-grained picture. It is noted that $\Delta_q$ encodes the spatial correlations of the effective structural pinning field. However, such pinning effect does not require the presence of the metallurgical disorder (e.g., solute atoms), *i.e.*, the interacting defects and interfacial configurations can provide different local environments even in a chemically-pure system [48-50]. During the (steady) post-crossover state with the nonzero mean velocity (*i.e.*, $v_0 > 0$), the instantaneous height can be decomposed into the mean migration and local fluctuation, *i.e.*, $y(x,t) = v_0 t + h(x,t)$. The quenched force is thus understood to be regularized over the correlation length $\ell_y$ along the migration direction. At fixed $\ell_y > 0$, and for fluctuations satisfying $|h(x,t)|/\ell_y \ll 1$, the Taylor expansion about the mean trajectory $y = v_0 t$ in the comoving *Lagrangian* reference frame yields [47, 51],

$$\eta_q(x, v_0 t + h) = \eta_q(x, v_0 t) + \left.\frac{\partial \eta_q}{\partial y}\right|_{y=v_0 t} h(x,t) + \mathcal{O}(h^2) \tag{40e}$$

where, the first term in the *r.h.s.* denotes the avalanche noise $\eta_{dis}(x,t) \equiv \eta_q(x, v_0 t)$. By evaluating its spatiotemporal correlation with **Eq**.(40c), we can arrive the motional averaging effect,

$$\langle \eta_{dis}(x,t)\eta_{dis}(x',t')\rangle = \Delta_q(|x-x'|)R_{\ell_y}\big(v_0(t-t')\big) \cong \frac{\Delta_q(|x-x'|)}{v_0}\delta(t-t') \tag{40f}$$

where, the prefactor $1/v_0$ follows from the normalization of the correlation kernel,

$$\int_{-\infty}^{\infty} R_{\ell_y}(v_0\tau)\mathrm{d}\tau = \frac{1}{v_0}, \quad v_0 > 0 \tag{40g}$$

where, $\tau = t - t'$ is the time interval. The temporal white-noise approximation in **Eq**.(40f) preserves the integrated covariance and applies on time scales much longer than the sampling correlation time $\ell_y / v_0$. It is noted that this sampling along the mean migration trajectory induces two consequences, *i.e.*, 1) under the above temporal coarse-graining condition, the quenched disorder undergoes motional averaging, emerging as an effectively annealed noise (white in time), while keeping its spatial correlation $\Delta_q(|x - x'|)$; 2) the integrated covariance is proportional to $1/v_0$, indicating a longer sampling correlation time at lower migration velocities.

Following the (MHKZ) RG framework established by Medina *et al*. [52], the noise spectrum with purely spatial correlation (equivalent to set the temporal correlation exponent $\theta = 0$ in Ref.[52]) can be defined in Fourier space as $\langle \eta_{dis}(q, \omega) \eta_{dis}(q', \omega') \rangle \propto |q|^{-2\rho} \delta(q + q') \delta(\omega + \omega')$. By performing the inverse Fourier transform to the temporal domain, we have the following correlation in $(q, t)$ space,

$$\langle \eta_{dis}(q, t) \eta_{dis}(q', t') \rangle \propto |q|^{-2\rho} \delta(q + q') \delta(t - t') \tag{40h}$$

where, $\rho > 0$ (is not the net defect density field defined in **§3.3.1**) defines the spatial correlation exponent characterizing the structural disorder [52].

Within the dynamic RG framework, the thermal noise has a wave-number-independent spectrum, whereas the spatially-correlated avalanche contribution scales as $|q|^{-2\rho}$ [53], which increases relative to the thermal background toward longer wavelengths for $\rho > 0$. In the pre-crossover regime ($v_0 \cong 0$), the periodic pinning term with the barrier of $\pi M \tilde{V}_0 / h_0$ (see **Eq**.(36)) suppresses the long-wavelength fluctuations. However, during the steady post-crossover migration, the periodic restoring contribution can be neglected under the temporal averaging conditions specified in **§3.3.6**, while the KPZ coefficient is still related to the reference migration velocity through the kinematics discussed in **§3.3.3**. In the post-crossover regime, the spatially-correlated noise associated with disconnection avalanches provides an additional source of interfacial fluctuations beyond the thermal background. As demonstrated by the MHKZ scaling laws, this disorder-dominated sliding state is characterized by the Hurst exponent $H = (1 + 2\rho)/3$, describing the roughened profiles and avalanche dynamics observed in MD simulations.

### *3.3.6. Transformation to the fractional Burgers equation*

To elucidate the nonlinear steepening associated with the morphological transition at the crossover point, we analyze the stability of the fKPZ interface against perturbations of the local slope. By taking the spatial derivative of the height profile, *i.e.*, $u(x,t) = \partial_x y(x,t)$, we transform the fKPZ equation **Eq.**(39) into the evolution equation for the local slope field $u$,

$$\frac{\partial u}{\partial t} = -\mu_{el}(-\Delta)^{1/2}u - M_\kappa \cos\left(\frac{2\pi y}{h_0}\right)u + \nu_2 \frac{\partial^2 u}{\partial x^2} - \nu_4 \frac{\partial^4 u}{\partial x^4} + \lambda u \frac{\partial u}{\partial x} + \frac{\partial \eta_{tot}}{\partial x} \tag{41}$$

where, $M_\kappa = 2\pi v_p/h_0 = 2\pi^2 M\tilde{V}_0/h_0^2$ is the local restoring coefficient obtained by differentiating the periodic pinning potential. In the pre-crossover regime, the interface is trapped near the potential valleys ($y \approx 0$, $h_0$, $2h_0$, ...), rendering $\cos(2\pi y/h_0) \approx 1$. The second term (in the *r.h.s.* of **Eq.**(41)) thus operates as the pinning barrier ($-M_\kappa u$) that localizes the interface and suppresses local slope fluctuations. Besides, the KPZ coupling $\lambda$ is negligible in the pre-crossover regime. Thus, the interface is shielded from nonlinear instabilities, maintaining the EW-like roughness.

Upon crossover, as the GB enters the steady-motion state with the finite velocity $v_0$ (*i.e.*, $y \approx v_0 t$), the periodic pinning term oscillates as $\cos(2\pi v_0 t/h_0)$. Under the sense of motional averaging, this oscillatory contribution averages to approximately zero over timescales exceeding the kinematic migration period ($t \gg h_0/v_0$), provided the slope field varies slowly over each period. Thus, the periodic restoring contribution (*i.e.*, the second term in the *r.h.s.* of **Eq.**(41), $-M_\kappa \cos\left(\frac{2\pi y}{h_0}\right)u$) can be neglected in the temporally coarse-grained description.

Simultaneously, the onset of GB migration induces the nonlinear KPZ coupling (recall $\lambda = 2v_0$ derived in **§3.3.3**). In the long-wavelength limit, the higher-order spatial derivatives ($\nu_2$, $\nu_4$) become subdominant. By neglecting the stochastic noise provisionally, the morphological dynamics reduces to the (deterministic) *fractional Burgers equation* [54],

$$\frac{\partial u}{\partial t} \cong -\mu_{el}(-\Delta)^{1/2}u + \lambda u \frac{\partial u}{\partial x} \tag{42}$$

Unlike conventional systems regulated by the *Laplacian* $\nabla^2$, **Eq.**(42) describes the competition between, the fractional elastic relaxation $(-\Delta)^{1/2}u$ and the advective term $u\partial_x u$. The nonlinear advection promotes the steepening of the slope field, while the fractional elasticity provides a competing relaxation mechanism [55]. For $\mu_{el} > 0$ and smooth periodic initial data, the *fractional Burgers equation* remains globally regular [54], however, this does not exclude finite amplification of the slope gradients or the occurrence of steep-front configurations [56].

Now, we consider the localized perturbation of the GB slope field characterized by the amplitude $u_0$ and spatial extent $L$. The characteristic time for the fractional elasticity to smooth out this perturbation thus scales as $t_{diss} \sim L/\mu_{el}$. Simultaneously, the nonlinear advection term drives the steepening of wavefronts, with the characteristic time of nonlinear steepening scaling as $t_{conv} \sim L/(\lambda u_0)$. It is noted that the length scale $L$ cancels out in the ratio of $t_{diss}$ and $t_{conv}$ (*i.e.*, $\lambda u_0/\mu_{el}$), rendering this dynamic competition scale-independent.

With the periodic restoring term averaged out, the condition $\lambda u_0 > \mu_{el}$ indicates that the characteristic time for nonlinear advection is shorter than that for (nonlocal) elastic relaxation, thus the nonlinear steepening of the GB slope field is favored. It is noted that the steepening here refers to finite amplification of local gradients, rather than an inevitable finite-time divergence of $|\partial_x u|$ (*i.e.*, $|\partial_x u| \to \infty$). Building upon the deterministic reduction of **Eq**.(42), the following **§3.4**. examines how spatially-correlated noise associated with disconnection avalanches contributes to the post-crossover roughening.

### 3.4. Avalanche dynamics in post-crossover regime

In the post-crossover regime, the interface continuously advances with a steady-state migration velocity $v_0 > 0$. As established in **§3.3.3**, this activates the KPZ nonlinearity ($\lambda = 2v_0 > 0$). The evolution of GB profile is therefore governed by $\partial_t y = v_0^e - \mu_{el}(-\Delta)^{1/2}y + \frac{\lambda}{2}(\partial_x y)^2 + \eta_{tot}(x,t)$. To evaluate the scale-invariant morphological roughness, the Galilean transformation is applied to the comoving frame, with denoting the fluctuation field as $h(x,t) = y(x,t) - v_0^e t$. The governing SPDE for the fluctuation field reads,

$$\frac{\partial h}{\partial t} = -\mu_{el}(-\Delta)^{1/2}h + \frac{\lambda}{2}\left(\frac{\partial h}{\partial x}\right)^2 + \eta_{tot}(x,t) \tag{43}$$

where, we omit the capillary and hyperviscous terms ($\nu_2\nabla^2 h$ and $-\nu_4\nabla^4 h$), as they are subleading in the long-wavelength limit ($q \to 0$).

#### *3.4.1. Derivation of the spatially-correlated noise spectrum*

As established in **§3.3.5**, the total noise field $\eta_{tot}(x,t)$ enters the power-law scaling regime upon the crossover. While the thermal noise $\eta_{th}(x,t)$ persists merely as the thermodynamic background, the GB migration is dominated by the collective behavior of disconnections, described by the spatially-correlated noise component $\eta_{dis}(x,t)$. To derive the correlation exponent $\epsilon$ of this avalanche-induced noise (where, $\epsilon$ parameterizes the spatial spectrum $S_\eta(q) \propto |q|^{-\epsilon}$, equal to $2\rho$ in the MHKZ theory), we evaluate the statistics of disconnection avalanches. Characterized by the fractional operator $-(-\Delta)^{1/2}y$ (*i.e.*, the $1/r$ nonlocal elasticity), a localized avalanche over a spatial extent $R$ is compared to a mode-II shear crack [57, 58]. According to the linear elastic fracture mechanics, such an avalanche is approximated as semi-elliptical spatially [57, 59],

$$s_R(x) = s_0\sqrt{1-(x/R)^2} \quad \text{for } |x| \leq R \tag{44}$$

where, the spatial Fourier transform of this semi-elliptical profile yields $\tilde{s}_R(q) \propto J_1(qR)/q$, where $J_1$ is the first-order Bessel function of the first kind [60]. Thus, the PSD of a single avalanche of size $R$ is $S_R(q) \propto J_1^2(qR)/q^2$, while the spatially-correlated noise spectrum $S_{\eta,dis}(q)$ experienced by the moving interface is the superposition of these avalanche profiles over all possible sizes $R$. At the crossover point, the avalanches exhibit a scale-free length distribution $P(R) \propto R^{-\tau_R}$. By integrating the individual spectral contributions, we can obtain the continuum spectrum as,

$$S_{\eta,dis}(q) \propto \int_0^{\infty} P(R) S_R(q) dR \propto \int_0^{\infty} R^{-\tau_R} \frac{J_1^2(qR)}{q^2} dR \tag{45a}$$

By introducing a change of variables $u = qR$ (with $\mathrm{d}R = \mathrm{d}u/q$), the above integral is recast as, $S_{\eta,dis}(q) \propto q^{\tau_R - 3} \int_0^{\infty} u^{-\tau_R} J_1^2(u) du$. Since the integral over $u$ yields a dimensionless constant, the spatially-correlated noise spectrum scales as $S_{\eta,dis}(q) \propto |q|^{\tau_R - 3}$. As established by the MHKZ framework $S_{\eta,dis}(q) \propto |q|^{-\epsilon}$, we derive the relation connecting the avalanche distribution to the noise spectrum as, $\epsilon = 3 - \tau_R$. In the long-wavelength limit ($q \to 0$), the total noise spectrum $S_{\eta,tot}(q) = S_{\eta,th}(q) + S_{\eta,dis}(q) \approx D_{th} + D_{dis}|q|^{-\epsilon}$ is dominated by the spatially-correlated contribution at small (but nonzero) wavenumbers ($|q|^{-\epsilon} \to \infty$), thus the thermal contribution becomes asymptotically irrelevant. It is noted that, the spectral amplitudes ($D_{th}$ and $D_{dis}$) are related to the thermal bath and motional averaging through **Eqs**.(40b) and (40f), respectively. For the thermal background, by evaluating the FDT-constrained correlator **Eq**.(40b), we arrive to the constant white-noise intensity $D_{th} = 2Mk_BT/L_z$. For the spatially-correlated noise driven by disconnection avalanches, transforming the motional-averaged correlator **Eq**.(40f) links the dynamic noise amplitude to the spectrum of the quenched disorder (*i.e.*, $\mathcal{F}[\Delta_q(x)] \equiv A_q|q|^{-\epsilon}$, where the material constant $A_q$ is the amplitude of the quenched disorder), yielding $D_{dis} = A_q/v_0$. Under the motional averaging approximation for steady migration, $D_{dis}$ characterizes the amplitude of the time-integrated covariance, whose inverse-velocity dependence reflects the longer sampling correlation time at lower migration velocities. This white-noise approximation thus requires the observation time much longer than $\ell_y/v_0$ (defined in **Eq**.(40c)) and cannot be extrapolated to $v_0 \to 0^+$ within the fixed observation time.

Furthermore, to analyze the morphological stability governed by the *fractional Burgers equation*, one has to transform the height profile $y$ to the slope field $u = \partial_x y$. Since the spatial differentiation $\partial_x$ introduces a $q^2$ multiplier to the power spectrum in Fourier space, the transformed noise spectrum $S_\zeta(q)$ governing the slope field evaluates to,

$$S_\zeta(q) = q^2 S_{\eta,tot}(q) = D_{th} q^2 + D_{dis} |q|^{2-\epsilon} \quad (45b)$$

where, the spectrum $S_\zeta(q)$ approaches to 0 in the long-wavelength limit ($q \to 0$) for $\epsilon < 2$. Since $\zeta(x,t) = \partial_x \eta_{tot}(x,t)$, its spatial integral vanishes under periodic boundary conditions.

To determine the avalanche length-distribution exponent $\tau_R$, we use for reference the (plastic) yielding transition within the statistical mechanics of plasticity and fracture, where the avalanche size distribution $P(s) \propto s^{-\tau}$ for interfaces governed by long-range elastic interactions is captured by mean-field theory with the universal exponent $\tau = 3/2$ [61], consistent with the MD results shown in **Appendix A**.

The avalanche size $s$ and its spatial length $R$ are related by the fractal scaling of the slip profile $s \sim R^{1+\zeta}$, where $\zeta$ is the roughness exponent of the pinned interface. As established by the one-loop dynamic RG framework [62], the $1/r$ nonlocal interactions give rise to the roughness exponent $\zeta = 1/3$ (noting the numerical result with high precision $\zeta = 0.388$ [63]). Based on the conservation of probability, $P(s)\mathrm{d}s = P(R)\mathrm{d}R$, we obtain the relation of the size and length exponents as, $\tau_R = (\tau - 1)(1 + \zeta) + 1 = 5/3$. Then, substituting this length exponent into $\epsilon = 3 - \tau_R$, yields, $\epsilon = 4/3$.

*3.4.2. Dynamic scaling analysis with spatially-correlated noise*

To formulate the spatio-temporal correlation of this spatially-correlated noise, we should further establish its temporal statistics. Relative to the slow evolution of GB profile, individual avalanches can be treated as instantaneous, localized bursts. On timescales long compared with individual bursts, we approximate the noise as temporally-white while retaining its spatial correlations, *i.e.*, the spatiotemporal correlation is rewritten as $\langle \eta_{dis}(x,t) \eta_{dis}(x',t') \rangle = C(|x - x'|) \delta(t - t')$.

Here, the noise is delta-correlated in time thus has the frequency-independent spectrum, while the time-translation invariance gives $\delta(\omega + \omega')$, and the Fourier transform of $C(|x - x'|)$ yields the power spectrum $S_{\eta,dis}(q)$ derived above. The noise spectrum is thus established as, $\langle \tilde{\eta}_{dis}(q,\omega) \tilde{\eta}_{dis}(q',\omega') \rangle \propto |q|^{-\epsilon} \delta(q + q') \delta(\omega + \omega')$, with which we further perform the scale transformation to determine the Hurst exponent $H$. By rescaling space, time, and height with an arbitrary factor $b > 1$, we have,

$$x \to bx, \quad t \to b^z t, \quad h \to b^H h \quad (46)$$

where, $z$ is the dynamic exponent. Under this transformation, by balancing the scaling dimension of the temporal derivative ($\partial_t h \sim b^{H-z}$) with that of the spatially-correlated noise amplitude ($\langle \eta_{dis}\eta_{dis}\rangle \sim b^{\epsilon-z-d}$, where, $d = 1$ is the dimensionality), we have,

$$2H - 2z = \epsilon - z - 1 \Longrightarrow 2H - z = \epsilon - 1 \tag{47}$$

As established in the MHKZ framework [52], since the tilt symmetry (Galilean invariance) of the KPZ equation prevents the renormalization of the coupling coefficient $\lambda$, we can obtain the relation between the dynamic and Hurst exponents as,

$$z + H = 2 \Longrightarrow z = 2 - H \tag{48}$$

By substituting **Eq.**(48) into the noise scaling balance **Eq.**(47), we have, $2H - (2 - H) = \epsilon - 1 \Longrightarrow 3H - 2 = \epsilon - 1 \Longrightarrow H = (1 + \epsilon)/3$, consistent with the nonlinear scaling result reported by Katzav [64] for the fKPZ equation with spatially-correlated, temporally-white noise. Then, substituting the spatially-correlated noise exponent $\epsilon = 4/3$ (**§3.4.1**) into the MHKZ framework yields, $H = 7/9 \approx 0.778$, consistent with the post-crossover value $H \approx 0.8$ observed in MD simulations. This agreement supports the proposed role of spatially-correlated disconnection activity in the fKPZ description of GB roughening.

## 4. Results and discussions

### 4.1. Numerical evaluations of the fKPZ equation

To recover the transition of the Hurst exponent observed in MD simulations, we numerically integrate the SPDE by implementing the pseudo-spectral method coupled with the exponential time differencing scheme, handling the fractional and fourth-order terms through their exponential propagators. As shown in **Fig. 7**, by solving the fKPZ equation with the spatially-correlated noise featuring the correlation exponent of $\epsilon = 1.4$ [52], the numerical results show that a plateau of $H \cong 0.8$ is reproduced upon the crossover, consistent with the MD results shown in **Fig. 2**. The jump of the Hurst exponent from the (pre-crossover) $H \cong 1/3$ to the (post-crossover) $H \cong 0.8$ plateau is triggered by the cooperative evolution of three operators in the fKPZ equation. It should be noted that prior to the crossover, the interface is trapped near the potential valleys ($y \cong 0$). We thus expand the nonlinear pinning force to the leading order in the numerical implementation, arriving to a harmonic restoring force, *i.e.*, $-M\frac{\pi\tilde{V}_0}{h_0}\sin\left(\frac{2\pi y}{h_0}\right) \approx -\left(\frac{2\pi^2 M\tilde{V}_0}{h_0^2}\right)y \equiv -M_\kappa y$. Specifically, for $t < t_c$, the boundary is pinned ($M_\kappa \propto v_p$), thus suppressing long-wavelength fluctuations, while the KPZ nonlinearity is frozen ($\lambda \cong 0$). After the crossover point $t = t_c$ (where, $t_c$ is the prescribed crossover time in numerical protocols), the harmonic restoring coefficient decreases ($M_\kappa \to 0$), while the KPZ coefficient approaches to its steady state value ($\lambda \propto v_0$) and the fractional relaxation coefficient decreases (*i.e.*, $\mu_{el}$ softens) according to the specified protocols detailed in **Supplementary Materials**. As demonstrated in **Fig. 7**, our theory combines the (reduced) harmonic term, KPZ nonlinearity, and spatially-correlated noise ($\eta_{dis}$) together, reproduces the post-crossover plateau $H \cong 0.8$ in MD simulations. It is noted that this numerical implementation examines the roughness response under the prescribed crossover protocol, but does not independently predict the crossover strain of the Ni bicrystal.

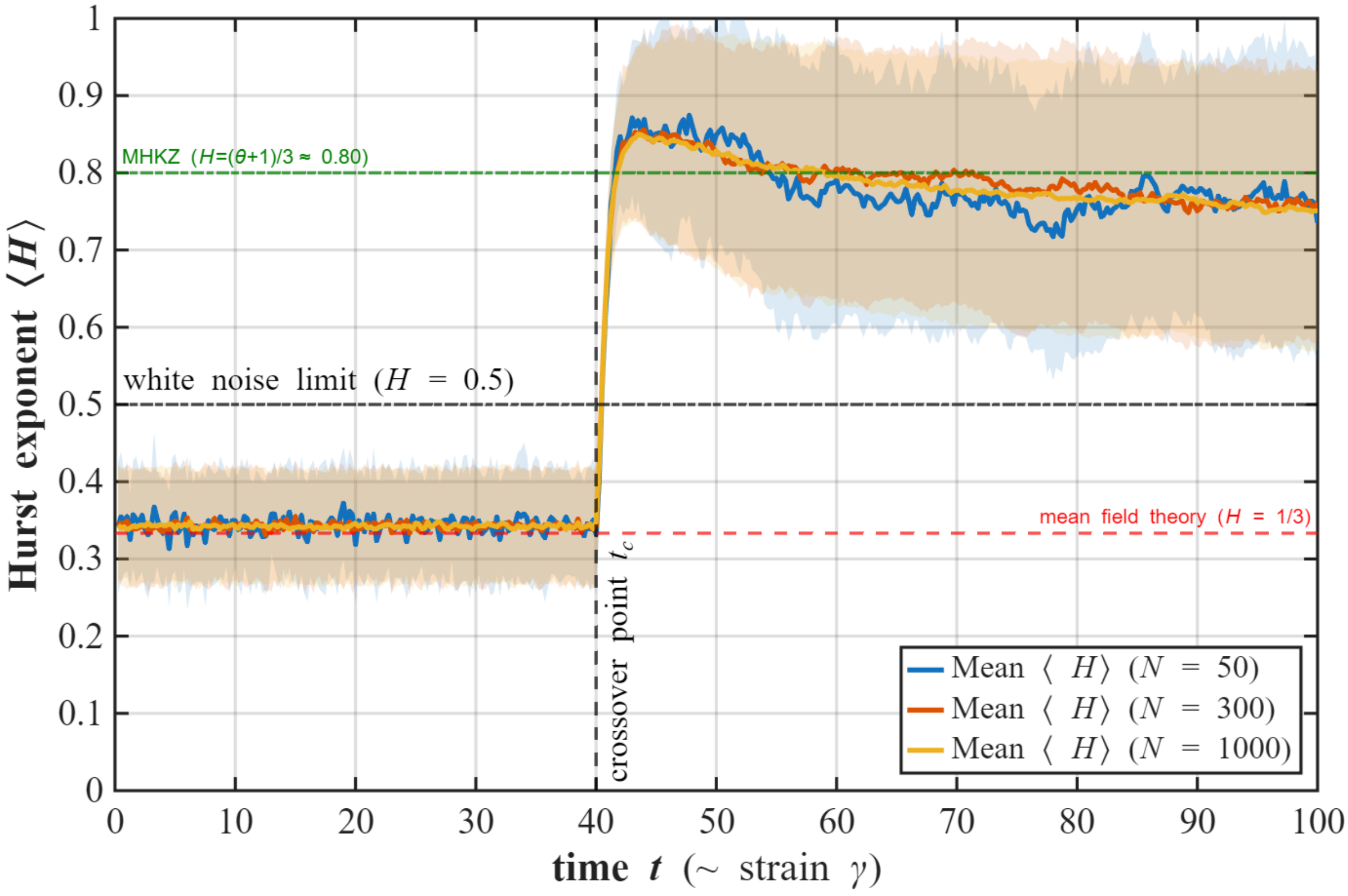

**Fig. 7**. Evolution of the Hurst exponent under the prescribed crossover protocol. The curves show ensemble averages of $N$ = 50, 300, and 1000 trajectories of the harmonic reduction of **Eq.**(39), with the correlation exponent of the spatially-correlated noise $\epsilon = 1.4$. The shaded bands indicate standard deviations across $N$ trajectories. The vertical dashed line marks the prescribed crossover time $t_c$. Time $t$ is expressed in model units.

### 4.2. Nonlocal elasticity and lattice pinning in pre-crossover regime

Prior to the crossover, the migration velocity of GB is negligible ($v_0 \approx 0$), thus deactivating the KPZ nonlinearity ($\lambda \approx 0$). In the long-wavelength limit ($q \rightarrow 0$), the higher-order damping terms $\nu_2 q^2$ and $\nu_4 q^4$ decay much faster than the elastic kernel $\mu_{el}|q|$, thus become subdominant. As established in the theory of elastic depinning [65], we neglect the thermal fluctuation $\eta_{th}(x,t)$ in the coarse-grained limit (in the pre-crossover regime) [51, 66]. By averaging out the pinning potential to isolate the interfacial relaxation, the fKPZ equation reduces to the qEW equation with nonlocal elasticity [63],

$$\frac{\partial y}{\partial t} = -\mu_{el}(-\Delta)^{1/2} y + \eta_q(x, y(x,t)) \tag{49}$$

While the coarse-grained pinning landscape can be treated as effectively quenched, this reduced **Eq**.(49) takes the form of the long-range qEW model, describing an elastic string subject to quenched pinning forces.

Herein, we clarify that the emergence of this 'random medium' does not require the existence of impurities or amorphous disorder, rather could arise even within a chemically pure bicrystal. In the continuum picture, the quenched disorder originates from the incommensurability and geometric frustration between the fluctuating interface and the *Peierls*-like potential. Specifically, the thermal roughening of GB introduces the stochastic distribution of local disconnection steps. When this roughened interface is superimposed onto the periodic potential, the local pinning phase becomes spatially randomized along the interface (*x*-axis direction). Since the height fluctuations far exceed the lattice periodicity, the spatial correlation of the pinning force decays significantly. Thus, at the coarse-grained level, we approximate the structural pinning by an effectively quenched disorder field $\eta_q(x, y(x,t))$ [67-69]. For the $1/r$ long-range elastic string in the random medium, the one-loop prediction is $H = 1/3$ [62], whereas the numerical calculations give $H = 0.388$ [63]. These results provide reference values for interpreting the pre-crossover roughness observed in present MD simulations and prior measurements on static GB fluctuations [1, 3], support the applicability of the effective quenched-disorder approximation.

Besides, this pre-crossover (pinned) landscape is sampled along the migration path, allowing the motional averaging after the onset of migration ($v_0 > 0$) as derived in **§3.3.5**. The rapid migration temporally smears out the (static) structural heterogeneities encountered along the migration path, dynamically projecting the quenched disorder $\eta_q(x, y(x,t))$ into the spatially-correlated noise $\eta_{dis}(x,t)$ that dominates the fKPZ scaling in post-crossover regime.

### 4.3. Morphological crossover during shear-coupled migration

As established in **§3**, the onset of morphological crossover indicates the change in the spectrum of interface fluctuations, *i.e.*, the quenched disorder gives way to the spatially-correlated noise $\eta_{dis}(x,t)$. In order to study the morphological stability in the post-crossover regime, we take the derivative of the fKPZ equation, $u = \partial_x y$, yielding the *fractional stochastic Burgers equation* (revisit **Eq.**(41) in **§3.3.6**),

$$\frac{\partial u}{\partial t} = -\mu_{el}(-\Delta)^{1/2}u + \lambda u\frac{\partial u}{\partial x} + \zeta(x,t) \tag{50}$$

where, $\zeta(x,t) = \partial_x[\eta_{th}(x,t) + \eta_{dis}(x,t)]$ represents the conserved noise derived as the spatial gradient of the dual-noise field. In Fourier space, this spatial differentiation introduces a $q^2$ multiplier to the noise power spectrum. Since the avalanche exponent $\epsilon = 4/3$ is less than 2, the transformed noise spectrum vanishes in the long-wavelength limit (*i.e.*, $\lim_{q\to 0} S_\zeta(q) \propto \lim_{q\to 0}[D_{th}q^2 + D_{dis}|q|^{2-\epsilon}]$) [52]. Besides, in the pre-crossover regime, the nonlocal elasticity suppresses the KPZ nonlinearity, preserving the sub-diffusive string profile. However, upon the crossover, the nonlinear steepening contends with the fractional elastic relaxation, as characterized by the ratio $\lambda u_0/\mu_{el}$. By taking the zero-noise limit, the nonlinear advection competes with fractional elastic relaxation and could produce (finite) local steepening of the GB slope field.

At the crossover point, the GB enters into the roughening regime dominated by disconnection avalanches. Evidenced by MD simulations, the Hurst exponent fluctuates at a plateau of $H \cong 0.8$ in the post-crossover regime. This post-crossover roughening breaks the classical $H = 0.5$ limit of the standard 1D KPZ equation driven by Gaussian white-noise.

**Eq.**(50) thus describes the competition between the nonlinear steepening and (nonlocal) elastic relaxation in the presence of stochastic force. The behavior of spatially-correlated disconnections is represented by the noise component $\eta_{dis}$, whose contribution to the roughness spectrum has been analyzed in **§3.4**. As established in **§3**, in the long-wavelength limit ($q \to 0$), the power spectrum of this dual-noise field is dominated by the spatially-correlated noise ($\eta_{dis}$), causing the thermal background ($\eta_{th}$) to be subleading.

Theoretically, within the dynamic RG framework for interface growth [52], the tilt symmetry of the KPZ nonlinearity, together with the noise scaling relation adopted in **§3.4.2**, gives $H = (1+\epsilon)/3$, where $\epsilon$ is the spatial correlation exponent of the spatially-correlated noise, $S_{\eta,dis}(q) \propto |q|^{-\epsilon}$. As derived in **§3**, the integration over the avalanche-length distribution gives, $\epsilon = 3 - \tau_R$. For interfaces governed by the $1/r$ nonlocal elasticity, by combining the mean-field size-distribution exponent $\tau = 3/2$ with $s \sim R^{1+\zeta}$ and $\zeta = 1/3$, the avalanche length-distribution exponent is established as $\tau_R = 5/3$ [65, 70], thus we have $\epsilon = 4/3$, and $H = 7/9$, which is in good agreement with the MD results of $H \approx 0.8$.

This post-crossover exponent $H \cong 0.8$ is also comparable to roughness exponents reported for dynamic fracture experiments [71]. While the *in-situ* tracking of 1D crack fronts remains experimentally elusive, the 1D out-of-plane trace of the crack surface, measured parallel to the crack front, could provide the indirect records of the crack front roughness [72, 73]. Specifically, in the quasi-static creep regime, the observed crack fronts exhibit $H \cong 0.35$ (consistent with the fluctuating line model) [74], comparable to the qEW prediction in the pre-crossover regime. However, during the dynamic failure driven by the coalescence of damages, analogous to the disconnection avalanches, the roughness exponents of 1D trace close to $H \cong 0.8$ have been reported for fracture surface profiles [73, 75-77].

This comparison thus highlights the collective behavior of defects (e.g., disconnections, dislocations) might contribute to roughening in both systems. Within the present model, the interplay between the nonlocal elasticity and spatially-correlated disconnection activity pushes the interface from the sub-diffusive state ($H \cong 0.33$) into the post-crossover roughening regime ($H \cong 0.8$).

## 5. Conclusions

In summary, we have formulated a theoretical framework that unifies the many-body dynamics of GB disconnections with the continuum field theory of kinetically-roughened interfaces. Our results reveal that the morphology of driven GBs undergoes the transition from a sub-diffusive elastic string to the (post-crossover) roughened fKPZ interface at low-temperature regimes (lower than $T_{KT}$ established in the ***CSH*** theory [15]).

In the pre-crossover regime, the GB morphology is governed by the $1/r$ nonlocal elasticity (operating as the *fractional Laplacian* $-(-\Delta)^{1/2}$), coupled with the geometric frustration imposed by the periodic lattice pinning, giving the Hurst exponent of $H \approx 0.33$, which is in quantitative agreement with MD results and previous measurements [1, 3, 63]. Upon the crossover, the interfacial morphology enters into the fKPZ regime. While the nonlinear steepening competes with (nonlocal) elastic relaxation, the collective motion of disconnections is represented by spatially-correlated noise. Under the adopted scaling assumptions, the present theory predicts $H = 7/9$, consistent with the observed roughness plateau of $H \cong 0.8$. The observed change in roughness scaling parallels that reported in dynamic fracture experiments, suggesting the common role for spatially-correlated avalanches in shaping the morphology of driven interfaces. Future efforts should be dedicated to the high-temperature regimes established in the ***CSH*** theory, where multiple-disconnection modes are activated.

**Acknowledgements**

This work was supported by the National Natural Science Foundation of China (Grant No. 12102145) and the Natural Science Foundation of Jiangsu Province (Grant No. BK20210444).

## Appendix A. The statistics of stress-drop events

In this section, the intermittent stress release is evaluated from the shear stress-strain curve of the 300 K simulation at the strain rate of $\dot{\gamma} = 10^7\ /s$. Following the event definition proposed in Refs. [38, 39], each maximal sequence of consecutive decreases in the recorded stress is identified as a stress-drop event, while a constant or increasing stress interval terminates the event. For the $j$-th event, starting at $\gamma_s^{(j)}$ ending at $\gamma_e^{(j)}$, the positive stress-drop amplitude is,

$$\Delta\tau^{(j)} = \tau\left(\gamma_s^{(j)}\right) - \tau\left(\gamma_e^{(j)}\right) > 0 \tag{A1}$$

Besides, the sampling window should exclude the initial elastic branch, where the disconnection avalanche is not activated. The sampled stress-drop amplitudes are then grouped into $M$ logarithmically-spaced bins spanning the observed amplitude range. Let $e_0, \cdots, e_M$ denote the boundaries of these $M$ bins (specifically, the bin $m$ spans $[e_{m-1}, e_m)$, with the upper boundary included in the final bin) and $n_m$ the number of events in the bin $m$, the probability density is estimated as,

$$\hat{p}(\Delta\tau_m^*) = \frac{n_m}{N_e(e_m - e_{m-1})}, \quad m = 1, \cdots, M \tag{A2}$$

where, $\Delta\tau_m^* = \sqrt{e_{m-1}e_m}$ is the geometric center of the bin $m$ used for plotting, and $N_e$ is the total number of stress-drop events included in the analysis. As shown in **Fig. A1**, if effective stiffness is approximately constant and the loading increment during an event is negligible, the stress-drop amplitude is proportional to the collective slip size, allowing the comparison with the mean-field reference $P(s) \propto s^{-3/2}$ (marked by the red dash line).

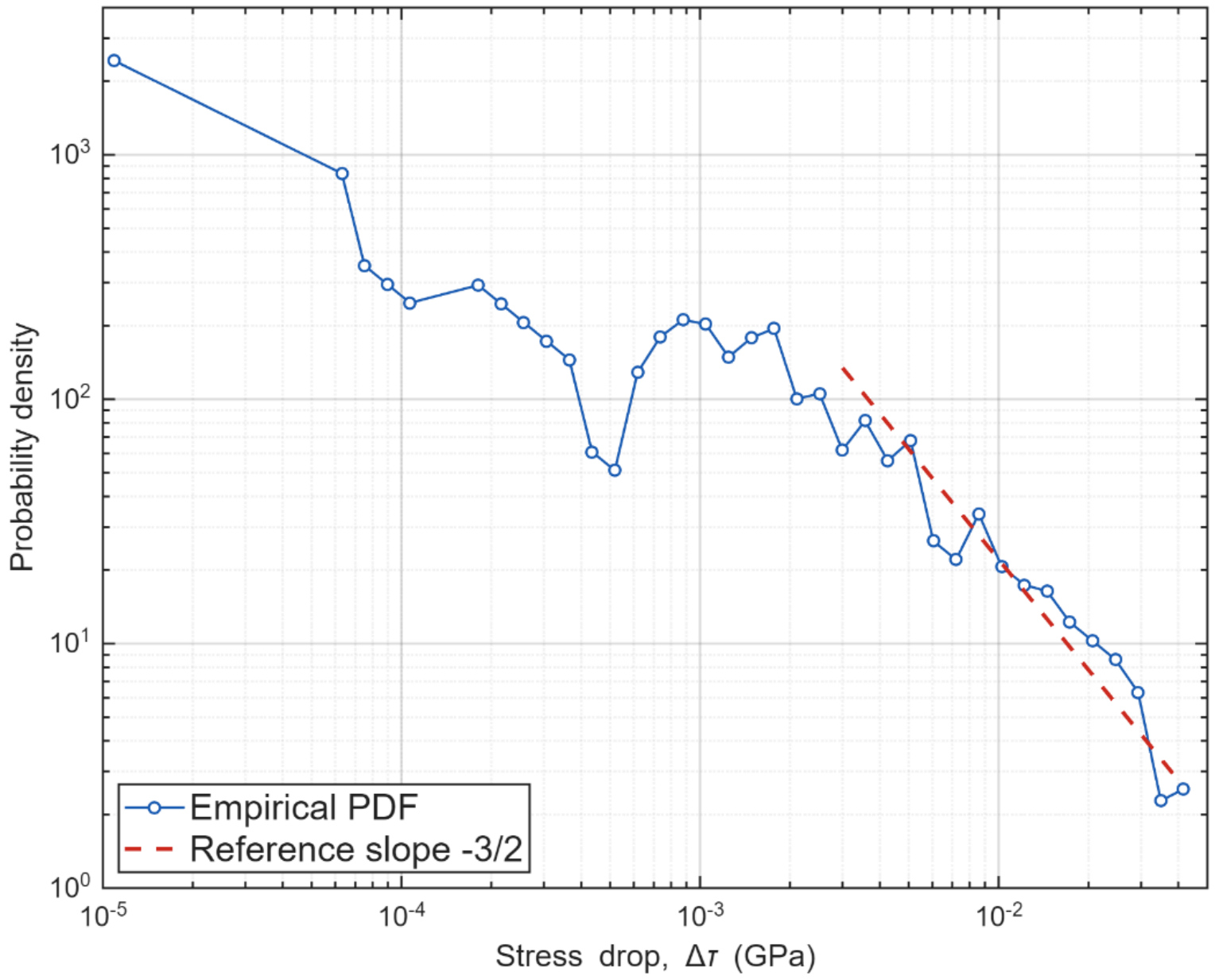

**Fig. A1**. Probability density of stress-drop events extracted from the 300 K curve shown in **Fig. 3**. The red dash line is not fitted from the data points, but the mean field reference line with the slope of -3/2.

## Appendix B. MEP of the GGBE landscape

By introducing the dimensionless coordinates $p = \phi/b_0$ and $q = y/h_0$, **Eq**.(1) can be rewritten as,

$$\mathcal{V}(p,q) = \frac{V_0}{2}[1 - \cos(2\pi p)] + \frac{U_0}{2}[1 - \cos(2\pi q)] + W_0[1 - \cos 2\pi(p - q)] \qquad \text{(B1)}$$

By using the Euclidean metric in these normalized coordinates, the **MEP** condition for a path $\boldsymbol{\mathcal{R}}(\alpha) = \big(p(\alpha), q(\alpha)\big)$ with unit tangent $\boldsymbol{t} = \boldsymbol{\mathcal{R}}_\alpha / |\boldsymbol{\mathcal{R}}_\alpha|$ ($\boldsymbol{\mathcal{R}}_\alpha \equiv d\boldsymbol{\mathcal{R}}/d\alpha = \left(\frac{dp}{d\alpha}, \frac{dq}{d\alpha}\right)$) is,

$$(\boldsymbol{I} - \boldsymbol{t}\boldsymbol{t}^T)\nabla_{p,q}\mathcal{V} = 0 \qquad \text{(B2)}$$

Thus, the potential gradient has no component normal to the path [40]. Here, the coordinates $p$, $q$ denote the local interfacial states, while the path parameter $\alpha$ is different from the spatial coordinate $x$ along the GB. For the shear-coupled path, a single period connects (0, 0) and (1, 1) in the $(p, q)$ space. By writing this path locally as $q = q(p)$, the tangential vector reads as,

$$\frac{d\mathcal{R}}{dp} = (1, q') \qquad \text{(B3)}$$

where, $q' = dq/dp$. Thus, the unit tangent $\boldsymbol{t}$ and corresponding unit normal $\boldsymbol{n}$ are,

$$\begin{cases} \boldsymbol{t} = (1, q')/\sqrt{1+q'^2} \\ \boldsymbol{n} = (-q', 1)/\sqrt{1+q'^2} \end{cases} \tag{B4}$$

respectively, and the GGBE gradient is $\nabla \mathcal{V} = (\mathcal{V}_p, \mathcal{V}_q)$, with,

$$\begin{cases} \mathcal{V}_p = \partial \mathcal{V}/\partial p = \pi V_0 \sin(2\pi p) + 2\pi W_0 \sin 2\pi(p-q) \\ \mathcal{V}_q = \partial \mathcal{V}/\partial q = \pi U_0 \sin(2\pi q) - 2\pi W_0 \sin 2\pi(p-q) \end{cases} \tag{B5}$$

Since **Eq**.(B2) requires the normal component of the GGBE gradient to be zero, *i.e.*,

$$\boldsymbol{n} \cdot \nabla \mathcal{V} = \frac{-q'\mathcal{V}_p + \mathcal{V}_q}{\sqrt{1+q'^2}} = 0 \tag{B6}$$

we have,

$$\mathcal{V}_q = q' \mathcal{V}_p \tag{B7}$$

According to the simplified string method developed by E *et al*. [78], the path is discretized into $N$ images in the normalized configuration space $(p, q)$, among which the *i*-th image at the *k*-th iteration is denoted as,

$$\boldsymbol{\mathcal{R}}_i^{(k)} = \left(p_i^{(k)}, q_i^{(k)}\right), \quad i = 0, \cdots, N-1 \tag{B8}$$

These images sample a path on the local GGBE surface, with the two endpoints fixed at the prescribed initial and final states, $\boldsymbol{\mathcal{R}}_A$ and $\boldsymbol{\mathcal{R}}_B$, respectively. Each iteration consists of the gradient relaxation followed by the redistribution of images along the updated path, which are detailed as follows,

*1) Gradient relaxation*. The interior images are first displaced along the full negative potential gradient,

$$\tilde{\boldsymbol{\mathcal{R}}}_i^{(k)} = \boldsymbol{\mathcal{R}}_i^{(k)} - \Delta\tau \nabla_{p,q} \mathcal{V}\left(\boldsymbol{\mathcal{R}}_i^{(k)}\right), \quad i = 1, \cdots, N-2 \tag{B9}$$

where, $\Delta\tau$ is the numerical relaxation step, unrelated to the physical time in the interface dynamics. The tilde denotes the intermediate positions before the redistribution. Both gradient components are evaluated at the current image coordinates,

$$\begin{cases} \tilde{p}_i^{(k)} = p_i^{(k)} - \Delta\tau \mathcal{V}_p\left(p_i^{(k)}, q_i^{(k)}\right) \\ \tilde{q}_i^{(k)} = q_i^{(k)} - \Delta\tau \mathcal{V}_q\left(p_i^{(k)}, q_i^{(k)}\right) \end{cases} \tag{B10}$$

with endpoints remain unchanged,

$$\tilde{\mathcal{R}}_0^{(k)} = \mathcal{R}_A$$
$$\tilde{\mathcal{R}}_{N-1}^{(k)} = \mathcal{R}_B \tag{B11}$$

*2) Arc-length redistribution*. While applying the gradient relaxation alone tends to concentrate images near the potential minima, to maintain approximately uniform spacing along the transition path, the images are redistributed after each relaxation step. We first calculate the cumulative polygonal arc-length of the intermediate string as,

$$\begin{cases} s_0^{(k)} = 0 \\ s_i^{(k)} = \sum_{j=1}^{i} \left| \tilde{\mathcal{R}}_j^{(k)} - \tilde{\mathcal{R}}_{j-1}^{(k)} \right|, \quad i = 1, \cdots, N-1 \end{cases} \tag{B12}$$

and define the normalized coordinates as,

$$\xi_i^{(k)} = \frac{s_i^{(k)}}{s_{N-1}^{(k)}}, \quad i = 0, \cdots, N-1 \tag{B13}$$

which serves as the interpolation nodes. By applying the piecewise cubic Hermite interpolation separately to the following two data sets,

$$\left\{ \left( \xi_i^{(k)}, \tilde{p}_i^{(k)} \right) \right\}_{i=0}^{N-1} \quad \& \quad \left\{ \left( \xi_i^{(k)}, \tilde{q}_i^{(k)} \right) \right\}_{i=0}^{N-1} \tag{B14}$$

we have the resulting (scalar) interpolants specifying the two coordinates of $\mathcal{L}^{(k)}(\xi)$ for any $\xi \in [0,1]$, which reproduces the image coordinates at the original interpolation nodes,

$$\mathcal{L}^{(k)} \left( \xi_i^{(k)} \right) = \left( \tilde{p}_i^{(k)}, \tilde{q}_i^{(k)} \right) = \tilde{\mathcal{R}}_i^{(k)}, \quad i = 0, \cdots, N-1 \tag{B15}$$

The images for the next iteration are obtained by evaluating this interpolant at equally-spaced normalized arc-length parameters $\alpha_i = i/(N-1)$,

$$\mathcal{R}_i^{(k+1)} = \mathcal{L}^{(k)}(\alpha_i), \quad i = 0, \cdots, N-1 \tag{B16}$$

Now, the above string-method-based calculation can be compared with the strong-coupling approximation established in **§3.1**. For the shear-coupled transition between (0,0) and (1,1), the analytical reference path is,

$$q = p, \quad y = \frac{h_0}{b_0} \phi \tag{B17}$$

As shown in **Fig. 4**, the resulting path follows the diagonal closely, while the numerical string is obtained by relaxing both coordinates under the full potential in **Eq.**(B1), without imposing this relation **Eq.**(B17). Besides, we also examine the transition path of a pure step between (0,0) and (0,1), for which the straight reference line $p = 0$ generally does not satisfy the **MEP** condition, since the potential gradient normal to this line is,

$$\mathcal{V}_p\big|_{p=0} = -2\pi W_0 \sin(2\pi q) \tag{B18}$$

which is nonzero away from the stationary points. The numerical string consequently bends away from $p = 0$ on either side of the saddle, as shown in **Fig. 4**, and passes through ($p$, $q$) = (0, 0.5), giving,

$$\Delta\mathcal{V}_{pure-step} = \mathcal{V}(0, 0.5) - \mathcal{V}(0, 0) = U_0 + 2W_0 \tag{B19}$$

while the numerically shear-coupled path passes through ($p$, $q$) = (0.5, 0.5), giving,

$$\Delta\mathcal{V}_{shear-coupled} = \mathcal{V}(0.5, 0.5) - \mathcal{V}(0, 0) = V_0 + U_0 \tag{B20}$$

consistent with the values numerically evaluated in **Fig. 4**b.